\documentclass[oldversion,fleqn]{aa}
\usepackage{natbib}
\bibpunct{(}{)}{;}{a}{}{,} 

\usepackage{graphicx}

\usepackage{txfonts}
\newcommand{\maxe}{MA$\chi$}
\newcommand{\fig}{Fig.~\ref}
\newcommand{\sect}{Sect.~\ref}
\newcommand{\teff}{$\mathrm{T_{eff}}$}

\newcommand{\feh}{[Fe/H]}
\newcommand{\tabl}{Table~\ref}
\newcommand{\fehteff}{[Fe/H] - \teff}
\newcommand{\degrees}{^{\mathrm{o}}}

\begin{document}
   \title{The age of the Milky Way halo stars from the Sloan Digital Sky Survey}


   \author{P. Jofr\'e
          \inst{1,2}
          \and A. Weiss\inst{1}}


   \offprints{P. Jofr\'e}
\institute{
   Max-Planck-Institut f\"ur Astrophysik, Karl-Schwarzschild-Str. 1, 85741 Garching, Germany \and Laboratoire  d'Astrophysique de Bordeaux,  2 rue de l'Observatoire, B.P.89 - 33 271 Floirac Cedex, France\\
  \email{jofre@obs.u-bordeaux1.fr}
  }

\authorrunning{Jofr\'e \& Weiss}
\titlerunning{The age of halo field stars}
   \date{}


  \abstract
{We determined  the age of the stellar content of the Galactic halo  by considering  main-sequence turn-off stars. From the large number of halo stars provided by Sloan Digital Sky Survey, we could  accurately detect the turn-off as a function of metallicity of the youngest dominant population, which was done by looking at the hottest (bluest) stars of a population.  Using the turn-off temperature  of a population of a given metallicity, we looked for the isochrones with that turn-off temperature and metallicity and found no age gradient as a function of metallicity.  This would mean that this dominating population of the Galactic halo formed rapidly, probably during the collapse of the proto-Galactic gas. Moreover, we could find a significant number of stars with hotter temperatures than the turn-off, which might be blue horizontal branch (BHB) stars, blue stragglers, or main sequence stars that are younger than the dominant population and were probably formed in external galaxies and accreted later on to our Milky Way. \\
Motivated by the current debate about the efficiency of gravitational settling  (atomic diffusion) in the interior of old solar-type stars, we used isochrones with and without settling to determine the ages.  When ignoring diffusion in the isochrones we obtained ages of 14-16 Gyr. This result is a strong argument against inhibited diffusion in old halo field stars, since it results in a serious conflict with the age of the Universe of 13.7 Gyr. The age obtained including diffusion in the isochrones was 10-12 Gyr, which agrees with the absolute age of the old globular clusters in the inner halo.
}{}


\maketitle


\section{Introduction}

The debate of how the Milky Way was formed is mainly based on two classic works. The first one,  by \cite{eggen62}, states that the Galactic halo formed during a rapid, monolithic collapse of the proto-Galactic gas cloud. The second one,  by \cite{searle_zinn},  claims that the halo was formed via accretion of small galaxies over several Gyr in a rather chaotic manner. These contrasting theories have been the subject of discussions during the past few decades \citep[see e.g. the reviews of ][]{majewsky93, helmi08}. The knowledge of accurate and absolute ages  for the various stellar populations plays an important role in finding the timescale for the formation of the Galaxy.

When determining the age of a star with isochrones, knowing its accurate chemical composition, effective temperature, surface gravity,  and distances is imperative for placing the star in a restricted region of the Hertzsprung-Russell diagram (HRD).  Stellar evolutionary models must also be accurate in order to date the star in a restricted time frame.

For halo stars, distances are generally poorly known, and determination of the stellar atmosphere parameters has associated errors as well. 
Uncertainties in the treatment of superadiabatic convection and gravitational settling of heavy elements can also influence the relation between the age of a star and its position in the HRD \citep{Chaboyer95, weiss02}.  Determining the age of individual stars is therefore also problematic. These ages are rarely known to an accuracy better than twenty percent \citep{gustafsson01}. This is especially true for old, distant stars in the Galactic halo, constituting a big obstacle in the exploration of early Galactic history.

 Until recently, globular clusters (GCs) were the main tracers of the Galactic halo, \citep[see e.g.][]{searle_zinn, chaboyer96, sarajedini97}.   
 This picture has changed in the past few years especially thanks all the stars collected by current stellar surveys. Examples are the Geneva-Copenhagen survey of the solar neighborhood \citep{nordstrom04}, the HK objective-prism \citep{beers92}, the Hamburg/ESO surveys \citep{christlieb01}, the Sloan Digital Sky Survey \citep[SDSS,][]{sdss}, and the Sloan Extension for Galactic Understanding and Exploration \citep[SEGUE,][]{SEGUE}. 

 The results of the analyses of these  surveys, combined with the higher  confidence in obtaining accurate GCs parameters,  has allowed a common conclusion to be reached for the formation of the Galactic halo.   For example, \cite{SW02}, \cite{de_angeli05}, and \cite{marin_franch09} using GCs, and \cite{SN06}, \cite{carollo07, carollo10} and \citet[and references therein]{SN10} using field  halo stars,  agree 
that  both \cite{eggen62} and \cite{searle_zinn} are partially right, i.e, one part of the Galactic halo (inner halo) was formed during the initial collapse, whereas the other part (outer halo) was formed slowly and chaotically via accretion.  However, the importance of the latter infall and accretion still needs to be quantified and the knowledge of absolute ages for the different stellar populations is still a missing piece in this puzzle. 

\begin{figure}[!ht]
\centering
\includegraphics[scale=0.45,angle=90]{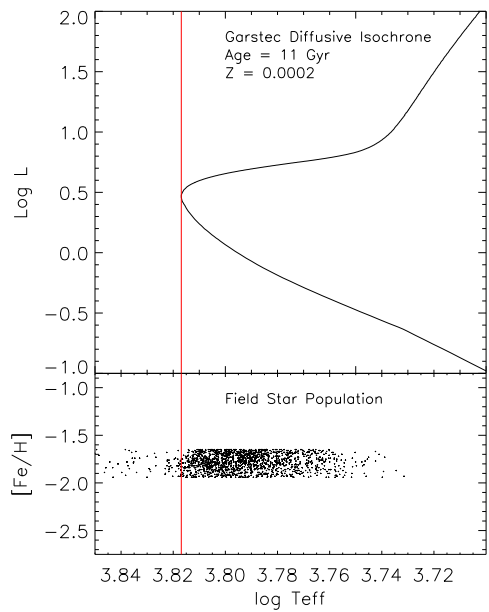}
\caption{Lower panel: Effective temperature of a SDSS field star population with metallicities in the range of $\mathrm{[Fe/H]} = -1.70 \pm 0.15$. Upper panel: 11~Gyr isochrone with metal fraction $\mathrm{Z} = 2 \times 10^{-4}$, which corresponds to the metallicity of the field population. The turn-off point of the isochrone is represented by the vertical red line, which agrees with the hottest stars of the field population. This agreement suggests an age 11 Gyr for the field population.}
\label{iso1D}
\end{figure}


In this paper we aim to obtain ages for the field stellar populations of the Galactic halo. 
Considering that we do not know the distance and mass of distant halo stars with enough accuracy, we can estimate ages of stellar populations by looking at the temperature and metallicity of the main-sequence turn-off. \\

\noindent In a dominating coeval stellar population of a certain metallicity, the turn-off point  (TO) corresponds to the stars with the  \textit{highest} effective temperature and the isochrone with this turn-off temperature provides the age of this stellar population, independently of the distance and mass of these stars. The only requirement   is to have a large sample of stars with independent measurements of metallicity and temperature, from where the hottest ones at a given metallicity can be found. 
 
This approach to estimating the age can be visualized in \fig{iso1D}, where the upper panel is an HRD with a theoretical GARSTEC isochrone  \citep{garstec08} of 11~Gyr and metallicity of $\mathrm{[Fe/H]} = -1.70$. The lower panel shows the temperature distribution of SDSS halo field stars  whose metallicities span the range of $\pm 0.15$ dex centered at the metal content of the isochrone. The turn-off temperature of the 11~Gyr isochrone from the upper panel  agrees with the hottest stars of the field population. This suggests an age of 11~Gyr for this  population. There is a significant number of stars hotter than this value. These stars can be BHB stars, stars of the same age but metal-poorer, blue stragglers, or younger stars of the same metallicity. They are present in all samples, contaminating the dominant population.  It is also important to mention that an older population, which would have a colder turn-off color than the dominating population, may well be hidden in this sample. Therefore, this method can only estimate the age of the youngest dominant population.

We are aware that this procedure is affected by further uncertainties. The effective temperature, for example, is difficult to obtain more accurately than within 100~K in the models. It depends on the atmosphere model and also on internal convection processes, which are uncertain. An error of 100~K in the temperature can produce an error of up to 2~Gyr in the age of population II stars.  By being aware of these uncertainties, we still can give an age estimate of the Milky Way halo and discuss implications ofor Galaxy formation.

 We first describe our sample and  the temperature and metallicity determination in \sect{atm_params}.  Second, we identify the main-sequence turn-off temperature in \sect{msto}, which we use to determine the age in \sect{age}.  Third, in \sect{gc} we  calibrate our results using the absolute ages of GCs and finally our conclusions are given in \sect{fin}.


\section{Sample and atmosphere parameters}\label{atm_params}

The stars  for this study  were taken from the SDSS/DR7 database \citep{dr7}, which also includes the SEGUE survey \citep{SEGUE}. We used a sample of 100,000 stars with low-resolution spectra, from which the metallicity and temperature can be estimated. This was done by performing a fitting between the observed spectra and a library of synthetic spectra. Because our sample of stars is large, a normal likelihood analysis can be very time-consuming. For that reason we used the \maxe~ method \citep{max} to optimize the fitting. 

The library of synthetic spectra was built using ATLAS12 \citep[Sbordone, 2010 priv. comm.]{atlas12_luca, atlas12_castelli} for generating the atmosphere models. They assume local thermodynamical equilibrium and a plane-parallel, line-blanketed model structure in one dimension. The grid of models cover a range of $-3.0 < \mathrm{[Fe/H]} < -0.5$ in steps of 0.1~dex, $5000 < \mathrm{T_{eff}} < 8000$~K in steps of 50~K and $3.5 < \log g < 5.0$ in steps of 0.5~dex. The synthetic spectra were then created with the SPECTRUM package \citep{spectrum} as described in \cite{max}, with the  difference that the input atmosphere models are now the ATLAS12 ones. The reason for this change in our grid of models was to avoid interpolations between the synthetic spectra, which allows more reliable final parameter distribution. Also, this new grid has larger stepsize in $\log g$, given that we were unable to determine $\log g$ with better accuracy than 0.5 dex in \cite{max} . Because our method of determining the turn-off  only depends on the temperature and the metallicity, we decided not to attempt to constrain the surface gravity parameter more than 0.5 dex. 

As in \cite{max}, we considered a wavelength range of [3900, 5500] \AA~ for estimating the metallicity and temperature.  It contains the Ca\small II K \normalsize absorption line at 3933 \AA, the Ca\small II \small H \normalsize line  at 3963 \AA, and the Mg\small I\normalsize b triplet at 5183 \AA, which we used as metallicity indicators. This wavelength range also includes the Balmer lines H$\beta$, H$\gamma$, and H$\delta$ at 4861, 4340, and 4101 \AA, respectively, which serve to estimate the temperature. The \maxe~method mainly  weights the spectra concentrating the fit on these features to determine the atmosphere parameters.  Since Ca and Mg are $\alpha-$elements,  a constant value of [$\alpha/\mathrm{Fe]} = +0.4$ was assumed in the synthetic spectra to obtain [Fe/H] from those metallic lines.   We are aware that this assumption might not be completely correct, as recently discussed by \cite{SN10},  who find an $\alpha-$poor (i.e. $\mathrm{[\alpha/Fe] = +0.2}$) and an $\alpha-$rich (i.e. $\mathrm{[\alpha/Fe] = +0.4}$) population in the halo. This would mean that some of the stars in our sample could have a 0.2 dex offset in metallicity, which is of similar to the accuracies obtained for metallicity with our method. Additionally, we do not expect a high contamination of the $\alpha-$ poor population found by \cite{SN10}, because it lies in the metal-rich regime of our sample. In this metallicity region we also have contamination of disk stars.

The mean error for the metallicities and temperatures obtained is of 0.25 dex and 140 K, respectively. These uncertainties are low enough to allow the determination of ages, as  discussed in \sect{age}.  We also considered the atmosphere parameters  provided by the SEGUE Stellar Parameter Pipeline \citep[SSPP,][]{ lee08_1} in order to check the consistency of our work.

We used color-color diagrams to select the F-G dwarfs and subgiants with  $0.1 < (g-r)_0 < 0.48$. This color constraint is based on the target selection for the $FG$  sample from SEGUE, which represents a metallicity unbiased random subsampling of subdwarfs \citep[see][for details on the SEGUE target selections from color contstraints]{SEGUE}.  The $FG$ sample, however, does not include the stars with $0.1 < (g-r)_0 < 0.2$, because a large number of BHB stars  have this color as well. For our purpose of determining the turn-off (i.e. bluest color) of each stellar population, we decided to include these blue stars, even when a high contamination with BHB is produced. We refer to this sample as the \textit{G blue} sample.

\section{Main-sequence turn-off}\label{msto}

Stars with a given metallicity value can have different temperatures distributed as seen in lower panel of Fig.\ref{iso1D}.  While  the number of stars increases smoothly at low temperatures, at high temperatures there is an abrupt decrease. In a dominating  stellar population we can interpret this drop  as the main-sequence turn-off  point (MSTO). In this section we describe the method employed to detect the turn-off and use metallicity-temperature diagrams to study the shape of the MSTO as a function of metallicity. 

\subsection{Turn-off detection: Sobel  Kernel edge-detection}\label{TO_TRGB}

For low-mass metal-poor stars, the  I magnitude of the tip red giant branch (TRGB) is almost constant.  \cite{Lee93}, \cite{Madore95}, \cite{Sakai96}, and \cite{Tabur09} have  used this knowledge to estimate distances of galaxies. Through histograms of the luminosity function, they determined the position of the TRGB by finding the luminosity where the count discontinuity is the greatest. They adopted a standard image-processing edge-detection technique to measure the magnitude of the tip, which is called Sobel Kernel. The main ingredient is a first derivative operator that computes the rate of change across an edge, where the largest change corresponds to the edge, e.g the tip.

In the same way that the TRGB shows an edge in the luminosity function of galaxies, the main-sequence turn-off  shows an edge in the temperature distribution function of stellar populations, and the Sobel Kernel can also be applied to this case.
In the following the edge-detection method is explained briefly only, but for further details see \cite{Lee93}, \cite{Sakai96}, \cite{Tabur09}, and references therein.

\begin{figure}[!ht]
\centering
\includegraphics[scale=0.45,angle=90]{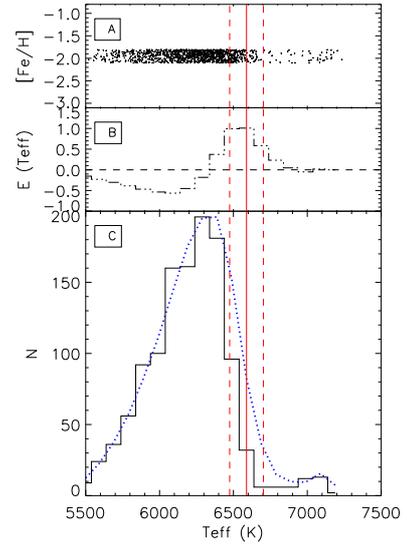}
\caption{Example of   Sobel Kernel  technique used to detect the turn-off. The TO is indicated by the red vertical lines, with the dashed lines corresponding to  their errors. Panel A:  Temperature distribution of the field stars with $\mathrm{[Fe/H]} = -1.95 \pm 0.15$. Panel B: Sobel Kernel filter response given by Eq.~(\ref{eq:ed}). Panel C: Histogram of temperature and the probability function of Eq.~(\ref{eq:phi}) by the blue dotted line.}
\label{to_techniques}
\end{figure}

We consider  a continuous probability distribution function to model the histograms. The dependencies of binning in temperature thus are avoided and the errors of the measurements are also included. The probability distribution can be defined by

\begin{equation}\label{eq:phi}
\Phi (\mathrm{T_{eff}}) = \sum_{i=0}^{N} \frac{1}{\sqrt{2 \pi}\sigma_i} \exp \left[ -\frac{(\mathrm{T_{eff}}-\mathrm{T_{eff,i}})^2}{2 \sigma_i^2} \right],
\end{equation}

\noindent where $\mathrm{T_{eff,i}}$, and $\sigma_i$ are the temperatures and their uncertainties, respectively, and $N$ the total number of stars in the sample.  The edge-detector filter applied to this temperature function is
\begin{equation}\label{eq:ed}
E(\mathrm{T_{eff}}) = \Phi (\mathrm{T_{eff}} - \Delta \mathrm{T_{eff}}) - \Phi (\mathrm{T_{eff}} + \Delta \mathrm{T_{eff}}),
\end{equation}
where $E(\mathrm{T_{eff}})$ is the filter response at the  temperature $\mathrm{T_{eff}}$, and $\Delta \mathrm{T_{eff}}$  the bin size.

Figure~\ref{to_techniques} illustrates  the technique. Panel A corresponds to the temperatures of the stars with $\mathrm{[Fe/H]} = -1.95 \pm 0.15$ form the SDSS survey.  To better visualize  the technique we plotted the temperature distribution with a histogram in  panel C  and the probability distribution  with the blue dotted line. The Sobel Kernel response given by Eq.~(\ref{eq:ed}) is in panel B, where the maximum response  is interpreted as the MSTO.

\subsubsection{Uncertainties in the turn-off detection}\label{errors}

\cite{Mendez02} and \cite{Tabur09} use the bootstrap method \citep{babu} to estimate uncertainties in the magnitude of the TRGB. Since our task is similar to the TRGB detection, we decided to use the bootstrap method to study the errors as described by these authors. We consider the uncertainty in our detection as $3 \sigma$, which corresponds to  99\% of the probability of the star being a turn-off star. The uncertainties for our sample with $\mathrm{[Fe/H]} = -1.95$ are plotted in Fig.~\ref{to_techniques} with dashed lines.

\subsubsection*{Effect of errors in metallicity measurements}

The effect of the metallicity measurement accuracies  of about 0.25 dex \citep{max} in the final TO-detection were analyzed with Monte Carlo simulations.   For a star with metallicity  value $\mathrm{\feh}_0$  and  error $ \sigma_{\mathrm{\feh}_0}$,  we gave a random value distributed as a Gaussian in the range of $[\mathrm{\feh}_0 - \sigma_{\mathrm{\feh}_0}, \mathrm{\feh}_0 + \sigma_{\mathrm{\feh}_0}]$.  This was done for each star, and then we looked for the turn-off in this new resample. As for the bootstrap error, we repeated this process 500 times and calculated the standard deviation of the TO-determination.


We found a significant   effect on the TO-detection errors due to the metallicity bin size used for the  temperature distribution.  This happens if the metallicity bin size is smaller than the averaged 2$ \sigma_{\mathrm{\feh}_0}$,  because a star can move from one bin to the other, which makes the temperature distribution change in those bins for that particular resampling. An effect in the TO-detection is produced, which is similar to the bootstrapping method explained above.

\begin{figure}[!ht]
\centering
\includegraphics[scale=0.4,angle=90]{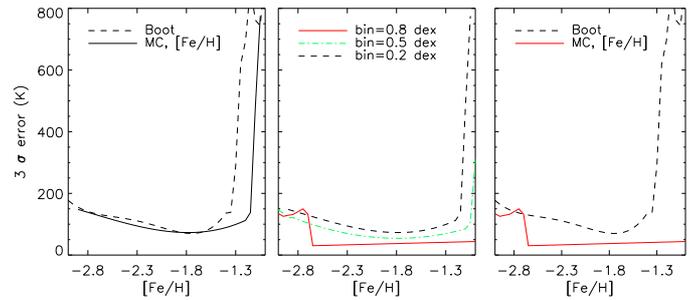}
\vspace{0.5cm}
\caption{Left panel: errors in the TO-determination due to bootstrapping method in temperature (Boot, dashed line) and due to Monte Carlo simulations of metallicity measurement (MC, [Fe/H], solid line) considering a metallicity bin size of 0.2 dex. Middle panel: errors in the TO-determination with Monte Carlo simulations using different metallicity bin sizes. Right panel: as left panel, but considering the bin size of 0.8 dex for Monte Carlo simulations.}
\label{errors_TO}
\end{figure}

In the left hand panel of \fig{errors_TO} we have plotted the error due to temperature bootstrapping (``Boot'') and the errors due to Monte Carlo (``MC'') simulations of metallicity measurements. The binning in the metallicity for the temperature distribution was  0.2 dex, which is less than the 2$ \sigma_{\mathrm{\feh}_0}$ accuracies of the metallicity measurements (0.25 dex).  We can see how both curves behave similarly, where the errors in the TO-detection become larger on the metal-rich side. This happens because there are fewer stars (discussed below) at high metallicities, meaning that a resampling of them will affect the shape of the temperature distribution more.

If the binning is much larger than the averaged 2$ \sigma_{\mathrm{\feh}_0}$ errors of the metallicity  measurements, there is less  probability that a star is moved from one bin to the other one, and the temperature distribution at that resampling remains unchanged.   This is shown in the middle panel of \fig{errors_TO}, where different curves represent the standard deviation of the TO-determination using different sizes in the metallicity binning. We can see how the errors become smaller with larger binning size, even at high metallicities, where there are fewer stars.  When the binning gets too large, the TO-detection becomes less accurate at the metal-poor border.    For a last comparison  we plotted the error due to temperature bootstrapping and  the error due to Monte Carlo simulations of metallicity measurements in the right hand panel of \fig{errors_TO}. In this case the binning in the metallicity for the temperature distribution was of 0.8 dex, which is much higher than the 2$ \sigma_{\mathrm{\feh}_0}$ errors of the metallicity measurements. The errors due to metallicity measurements become negligible when compared with the bootstrapping ones.  We must take into account that the cut-off at the turn-off becomes inaccurate if the metallicity bin is too large.  A  binning size of $\sim 0.5$ dex is the limit where we obtain a precise turn-off determination and where the errors due to metallicity measurements are negligible compared to those obtained with bootstrapping.

\begin{figure}[!ht]
\begin{center}
\includegraphics[angle=90, scale=0.45]{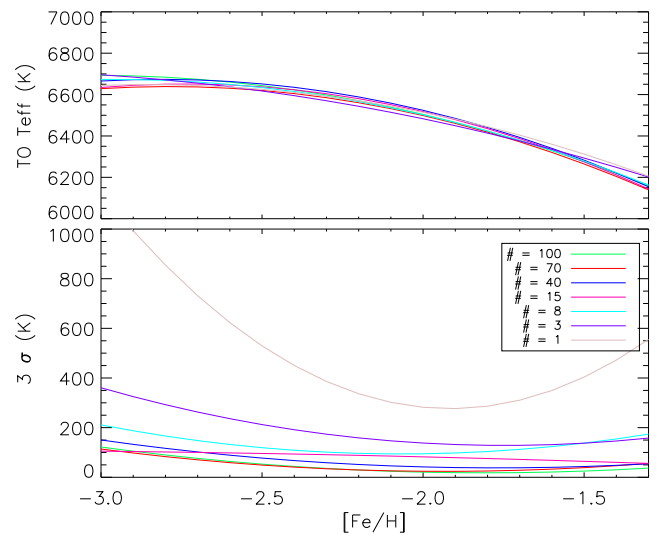}
\caption{Upper panel: TO-detection for samples with different number of stars. Lower panel: the bootstrap error in the TO-detection for each sample.  The color lines correspond to different sizes used for the detection, with the number of stars ($\times 1000$) indicated in the legend.  }
\label{err_num}
\end{center}
\end{figure}

\subsubsection{Uncertainties due to number of stars}

 For the stars of the \textit{G blue} sample, the larger their number, the smaller  the errors become in the TO-detection. This is shown in \fig{err_num}, where the curves represent different numbers of randomly selected stars from the \textit{G blue} sample. The amount of stars varies from 1,000 to 100,000. We plotted in the upper panel the value of the MSTO as a function of metallicity and  the errors obtained with bootstrapping in the lower panel . The TO-value does not depend on the amount of stars used in the sample, but there is a  dependency in the accuracies obtained. For the smallest sample the errors in the detection have the highest value.

\begin{figure}[!ht]
\begin{center}
\includegraphics[angle=90, scale=0.6]{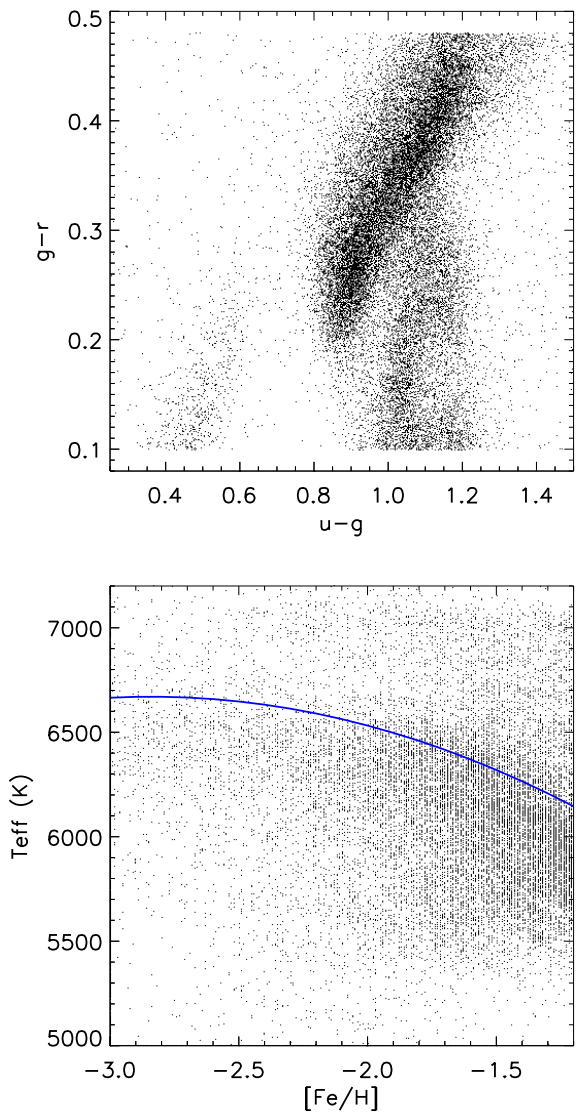}
\caption{Top panel: color-color diagram for the \textit{G blue} sample. Bottom panel: \fehteff~ diagram for the \textit{G blue} sample.  The turn-off temperature as a function of metallicity is plotted with blue color.}
\label{gblue}
\end{center}
\end{figure}

\subsection{Metallicity - Temperature diagram}\label{ft_diag}

The relation of the metallicity to the temperature of the stars can be confined in a metallicity - temperature (\fehteff) diagram.   Figure~\ref{gblue} shows an example of the diagram in the bottom panel. The top panel is the color-color diagram of the stars, which satisfies the color constraint described in \sect{atm_params}.  The majority of the stars are located in the main stellar locus in the color-color diagram, but we also see a second locus at blue $g-r$ colors, which corresponds to the BHB stars . They are  also present in the \fehteff~diagram at high  temperatures, but are not numerous enough to modify the Sobel Kernel edge detection.

In the \fehteff~ diagram  in the lower panel of \fig{gblue} we represent  the cut-off of the hottest stars. It is interpreted as the main sequence turn-off, and we determined it using the Sobel Kernel edge-detector technique explained above. At low temperatures  the number of stars gradually decreases. This is partly  due to selection effects, because our color constraint favors  dwarfs, which are located in the blue part of the main stellar locus. Another reason is evolution effects. Red giants evolve fast, meaning that we do not observe as many of them in the sky as main sequence stars. In addition, cold stars can also be early main-sequence stars, which are very faint and therefore  more difficult to observe than main-sequence stars close to the turn-off. 

The largest stellar population of our sample is the one at $\mathrm{\feh} = -1.65$, corresponding to the peak of the metallicity distribution of halo stars \citep{BeersChristlieb05}. At higher metallicities the turn-off halo stars are contaminated by  disk stars. Towards  lower metallicities, the number of stars decreases considerably, but they still populate the turn-off enough to employ the Sobel Kernel edge-detector. In the whole metallicity domain, there is a contamination of hot stars that can be BHB stars, white dwarfs, blue stragglers, or younger main sequence stars \citep{BMP, preston00}. They are not numerous enough to produce an effect in the turn-off detection, but further analyses of them are important for understanding  the halo structure and formation as a whole.

The final values of the turn-off temperature as a function of metallicity are listed in \tabl{tab_msto}. The first column represents the metallicities, in steps of 0.25 dex. For each step, we selected a metallicity window of bin size of 0.5 dex  to determine the edge in the temperature distribution applying  the Sobel Kernel edge-detector. At the second column of \tabl{tab_msto}, we listed the temperature at the turn-off obtained  for the \maxe~ parameters. The third column corresponds to the bootstrap errors in the TO detection, and the last two columns are the results of the age for the stellar populations at each metallicity. Isochrones with and without atomic diffusion were considered in the age determination (see \sect{age}).  For the errors in the age calculation, the uncertainties come from the bootstrapping of the turn-off detection and the consideration of 0.25 dex error in the metallicity measurement. The size of the metallicity bin used is of 0.5 dex, see \fig{errors_TO}.

\begin{table*}[ht!]
\begin{center}
\begin{tabular}{c c c c c}
\hline
\feh & $\mathrm{T_{eff,TO}}$ (K) & $\sigma_{\mathrm{TO}}$ (K) & Age$_{\mathrm{diff}}$ (Gyr) & Age$_{\mathrm{nodiff}}$ (Gyr)\\
\hline
-3.00 & 6691 &   146 &$12.0\pm1.73$&$17.0\pm 2.00$ \\
-2.75 & 6662 &    60 & $ 12.0\pm1.00$&$17.0\pm1.41$ \\
-2.50 & 6626 &    11  &$11.0\pm0.25$&$15.5\pm0.79$\\
-2.25 & 6582 &    3 & $11.5\pm0.25$&$16.0\pm0.25$ \\
-2.00 & 6531 &    19  &$11.5\pm0.71$&$16.0\pm0.93$\\
-1.75 & 6472 &    73  &$11.5\pm1.54$&$16.5\pm1.66$\\
-1.50 & 6405 &   172  &$11.0\pm2.31$&$16.0\pm2.31$\\
-1.25 & 6331 &   302 &$11.0\pm3.16$&$15.5\pm3.22$ \\
\hline
\end{tabular}
\end{center}
\caption{Final value of the turn-off temperature given by the field halo stars of SDSS at different metallicities.}
\label{tab_msto}
\end{table*}%

\subsubsection{Comparison with SSPP}

As a first application of the \fehteff~ diagram, we compared the distributions of the parameters recovered by using \maxe~\citep{max} and those obtained by the SSPP\citep[SSPP,][]{lee08_1, lee08_2, SSPPIII}. The comparison between the \maxe~ and SSPP parameters was discussed in  \cite{max}, where we showed the negligible general offset in effective temperature of 61 K for stars with temperatures between 5000 and 8000 K.  We also showed  a considerable offset of more than $\sim 0.3$ dex in metallicity for stars with $-0.5 < \mathrm{\feh} < -3.0$.
 
\begin{figure*}[!ht]
\hspace{1cm}
\includegraphics[scale=0.6, angle=90]{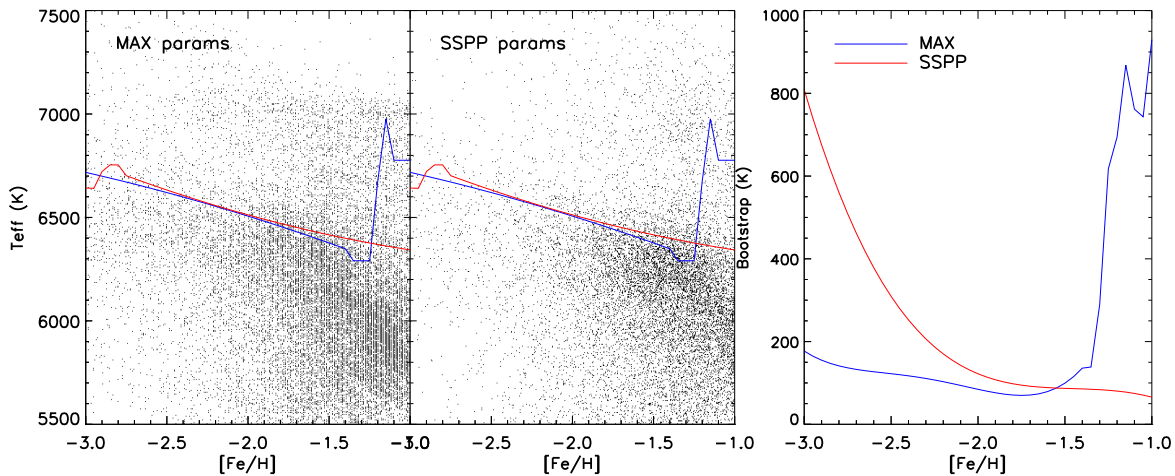}
\caption{Left panel: Temperatures and metallicities of the \textit{G blue} stellar sample (see \sect{atm_params}) for the atmosphere parameters estimated with \maxe. Middle panel: As left panel, but with the adopted SSPP \citep{lee08_1} atmosphere parameters. Blue line corresponds to MSTO temperature for the \maxe~ parameters and red line to SSPP ones. Right panel: Bootstrap errors in the TO detection as a function of metallicity.  }
\label{fehteff1}
\end{figure*}

The \fehteff~ diagrams for both cases are illustrated in \fig{fehteff1}, where in the left hand panel we find the \maxe~parameters and at the middle panel the SSPP ones. We looked for the turn-off using the Sobel Kernel edge detector and  compared the MSTO temperature as a function of metallicity. The $3\sigma$ bootstrap errors in the edge detection for both samples are compared in the right hand panel of the figure. 

 The differences due to the offset in metallicity can be seen in the right and middle panels of \fig{fehteff1} in the stellar distribution along the \fehteff~ diagrams. In the metal-rich regime we see  quite a large amount of stars for the SSPP parameters, while we do not for the \maxe~parameters. In a similar way, the metal-poor regime is more populated for the \maxe~ parameters than for the SSPP ones. 
It is worth noticing that the offset of $\sim 0.3$ dex in metallicity does not mean that every star is shifted by this value in metallicity. We can see in \fig{fehteff1} how the peak of the distribution is slightly  shifted, but the shape of the distribution is different for \maxe~ and SSPP.  This is seen especially  at low metallicities, where the \feh~ values between \maxe~ and SSPP parameters deviate the most.  We recall the recent comment  by \cite{carollo10} about the low \feh~ estimates of SSPP, which are 0.2 - 0.3 dex higher than those values recovered by high-resolution analyses. This would mean that the \maxe~ parameters at the metal-poor regime yield consistent results, which was also shown in \cite{max} in the test with high-resolution UVES spectra. 

In general, both distributions do not yield significantly different values of the turn-off  temperature for $\mathrm{\feh} < -1.2$. The lack of \maxe~ metal-rich stars does not allow any detection of an edge for $\mathrm{\feh} > -1.2$. In a similar way, the lack of SSPP metal-poor stars  affects the determination of the TO at  $\mathrm{\feh} < -2.7$.   The bootstrap errors  become larger at low metallicities for the SSPP parameters and at high metallicities for the \maxe~ parameters. This is because the different numbers of stars seen in the metal-poor and metal-rich regimes for SSPP and \maxe~ parameters. More stars yield a more accurate edge-detection (see \sect{errors}). 

The errors of the TO-detection for the SSPP parameters decrease smoothly with increasing metallicity, which agrees with the shape of the parameter distribution in the \fehteff~ diagram of the middle panel in \fig{fehteff1}, where the number of stars increases gradually towards higher metallicities. The errors of the TO-detection for the \maxe~ parameters, however, are relatively small for metallicities below $-1.4$, the point at which the errors increase abruptly towards higher \feh. This can be seen in the left hand panel of \fig{fehteff1}, where  the stars disappear abruptly at high metallicities. These differences do not introduce significant effects when determining the ages, as is discussed in \sect{age}.

 \begin{figure}[!ht]
\centering
\includegraphics[scale=0.5,angle=90]{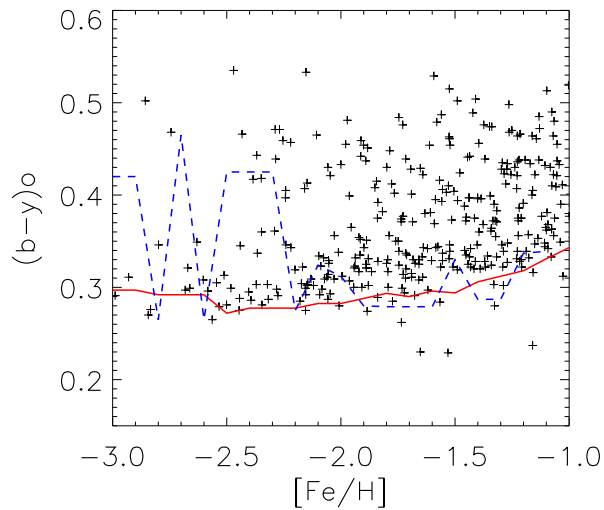}
\caption{$(b-y) - \mathrm{\feh}$ diagram of the photometric sample of Schuster et al.  (Salaris, priv. comm.). The red line corresponds to the MSTO determined from the SDSS sample as indicated in \tabl{tab_msto} and transformed to Str\"omgren photometry using the empirical relations of \cite{ramirez05}. There is a perfect agreement between the cut-off of SDSS and the  $uvby-\beta$ samples. The blue dashed line corresponds to the MSTO determined applying the Sobel Kernel edge-detector to the stars of the $uvby-\beta$ sample.}
\label{by_cols} 
\end{figure}

\subsubsection{Comparison with photometric sample}\label{comp_fehteff}

The sequence of works called ``$uvby-\beta$ photometry of high-velocity and metal-poor stars'' of W. Schuster, P. Nissen and collaborators  analyzes a sample of halo-field stars \citep[e.g.][]{SN88, SN06}. This series consists of several papers about kinematics, chemistry, and ages of halo-field stars. We considered the sample of \citet[hereafter called   $uvby-\beta$ sample]{SN06}, together with our SDSS \textit{G blue} sample to show that the MSTO temperature as a function of metallicity is unbiased. 

Our stellar sample has  $ugriz$ colors and the $uvby-\beta$ one has Str\"omgren photometry, and  therefore we had to transform colors to join both samples.  We only considered the MSTO temperatures  indicated in Table~\ref{tab_msto} and transformed them to $(b-y)$ colors using the scale of \cite{ramirez05} for dwarfs.   Because our temperature values are very similar over the entire metallicity range, we use the fine grid version of the tables of  I. Ram\' irez (priv. comm.), where the steps in temperature have a size of 50 K.   

The resulting MSTO $(b-y)$ color as a function of metallicity is plotted in \fig{by_cols}, where we also placed the colors and metallicities of the $uvby-\beta$ sample (Salaris, priv. comm.). The turn-off detected with spectroscopic temperatures in the SDSS survey agrees very well with the blue cut-off of the $uvby-\beta$ sample. 

We would like to point out the advantages of using the SDSS survey.  The sample is huge, which means that the MSTO as a function of \feh~ is  accurate. As a test, we applied the Sobel Kernel edge-detector (see \sect{TO_TRGB})  to find the $(b-y)$ color cut-off traced by the turn-off in the $uvby-\beta$ sample.  The values obtained as a function of metallicity are illustrated in \fig{by_cols}. The $uvby-\beta$ sample contains very few stars for some \feh~ values (especially those at low metallicities) and the turn-off cannot be detected  properly by the Sobel Kernel. For that  same \feh~ value, however, the SDSS data provide a clear cut-off and the Sobel Kernel yields realistic TO values.  

Given the confidence obtained in the studies  about field halo stars from the  $uvby-\beta$ sample, it is encouraging to obtain this excellent agreement in the blue cut off. Considering that both stellar samples are independent and that the analysis of them was performed using completely different techniques, this agreement implies that we have found the turn-off temperature as a function of metallicity of the Milky Way halo, independently  of the use of the SDSS spectroscopic or  $uvby-\beta$ photometric data set. These implications allow us to use the MSTO values in \tabl{tab_msto} in the following parts of this work because they are unbiased.

\section{Age determination}\label{age}

When determining the age of a stellar population based only on its metallicity and turn-off temperature, the effects of gravitational settling of heavy elements (also called atomic diffusion) can produce a considerable difference in the results \citep{chaboyer92a, chaboyer92b, salaris00}.  Owing to  settling, the content of heavy elements in the surface progressively decreases during the main-sequence phase as it sinks below the convective envelope.  Hydrogen is thus displaced out of the center towards the surface. The less hydrogen there is to burn in the core, the  faster  the evolution of the star along the main-sequence. This effect is particularly important in old metal-poor stars, because  they have shallow convective envelopes and are exposed to diffusion for longer times.

Although  atomic diffusion is a basic physical effect, its effectiveness
in stars is still being debated. In the solar case, a satisfying agreement  
needs to be  achieved  between a solar model and the
seismic sun \citep[e.g.][]{bp:95}. In low-mass, metal-poor globular
cluster stars there is some spectroscopic evidence of its presence
\citep[e.g.][for the case of NGC~6397]{korn06,lind08}, but this evidence
has been disputed \citep{gratton01}. Besides the change in the surface abundances
over time (which implies that the presently observed metallicity has
not been the initial one), gravitational settling of heavy elements affects the surface temperature
in two ways: directly through the composition-dependent opacity in the
outermost layers and indirectly through the accelerated evolution of the
stellar core, which tends to be more compact than when diffusion is not
taking place. As a result, $T_\mathrm{eff}$ is lower in stars where
diffusive settling is taking place. These two effects have been studied by \citet{chaboyer01} by
simultaneously including the age-reduction effect for
GCs in the isochrones   and  by inhibiting diffusion in the outermost stellar layers to prevent iron depletion from the surface. The latter mechanism  could be
due to either ``turbulent diffusion'' \citep{korn06}, ``rotation-induced
mixing'' \citep{mixing_rotation}, or some other effect.

While in the case of GCs, age indicators can be used that only
depend  very weakly on diffusion \citep{chaboyer92a, Meissner06},
for field stars these
effects become crucial when 
$\mathrm{T_\mathrm{eff}}$ (or the color) at the turn-off is being used,  as in the present case. If gravitational settling is ignored in the isochrones employed \citep[i.e][]{BV92}, the absolute ages obtained for the field stars can be up to 18~Gyr \citep{SN96}. Similarly, \cite{unavane96} obtained ages of 15-16~Gyr for a stellar sample of \cite{carney94} using \cite{Yale87} isochrones. These ages conflict with the age of the Universe \citep[13.7~Gyr,][]{wmap}. Stellar evolutionary models have improved over the years not only because of  considering atomic diffusion, but also better handling of opacities and $\alpha$-enhanced chemical compositions. This has also significantly decreased the absolute ages of old stars \citep{Chaboyer95, SW97}.  Moreover,  color transformations for the isochrones can also affect the absolute ages based on the turn-off color.   An example can be found in the  $uvby-\beta$ series, where \cite{SN06} found an age of 13~Gyr for the halo stars using newer isochrones of \cite{BV01}, which ignore diffusion but use empirical color transformations that require in some cases ad hoc adjustments to lower the ages based on a good match between the isochrones and the GC  turn-off colors. 

\cite{salaris_idea}  have obtained field star ages lower by up to 4~Gyr when using stellar
model tracks that included diffusion. Since that work was hampered by
having few objects, \cite{salaris_idea} suggest using
SDSS data to investigate this question with a larger statistical
sample. 
In the first part of this project \citep{max}, we analyzed the SDSS
stellar spectra to obtain metallicities and temperatures of
halo field subdwarfs. This is the necessary prerequisite for age
determinations.  The present work builds on this with the aim of 
determining the age of SDSS turn-off stars, for both cases of
including or ignoring atomic diffusion the stellar models.

\begin{figure}[!ht]
\centering
\vspace{-0.5cm}
\includegraphics[scale=0.45,angle=90]{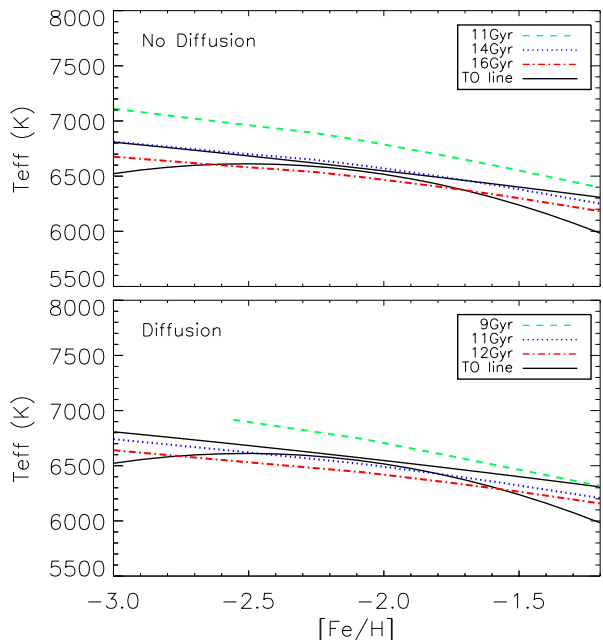}
\caption{Solid line: Turn-off  temperature range as a function of metallicity of the \textit{Gblue} sample. Color lines:  turn-off temperature and metallicity of GARSTEC isochrones. Upper panel: nodiff-isochrones of 11, 14, 16 Gyr reproduce the MSTO  temperature of the field stars. Lower panel: diff-isochrones of 9, 11,12 Gyr reproduce the MSTO temperature of the field.}
\label{fehteff_iso}
\end{figure}

\subsection{Age reduction due to atomic diffusion}\label{age_red}

Following the method employed by  \citet[and references therein]{SN06} and \cite{unavane96}  we looked for the turn-off temperature and metallicity of the isochrones that coincide with the cut-off traced by the main sequence turn-off in the \fehteff~ diagram. We considered GARSTEC isochrones  \citep[Z. Magic, priv. comm. 2010]{garstec08} with and without diffusion. Hereafter we refer to isochrones with diffusion as ``diff-isochrones'' and isochrones without diffusion as ``nodiff-isochrones''.  Settling of helium is responsible for the major change in effective temperature at the turn-off. Diffusion of other heavier elements produce a little additional change to what  is produced by helium \citep{weiss00_dif, chaboyer07}.  GARSTEC models consider full diffusion of helium and hydrogen using the diffusion coefficients of \cite{thoul94}, with no consideration of radiative levitation or rotational mixing.  Since we are assuming that most of the stars have a chemical composition that is $\alpha-$enhanced, we used isochrones according to this composition, as described in \cite{Meissner06}.

 \begin{figure}[!ht]
\centering
\includegraphics[scale=0.45,angle=90]{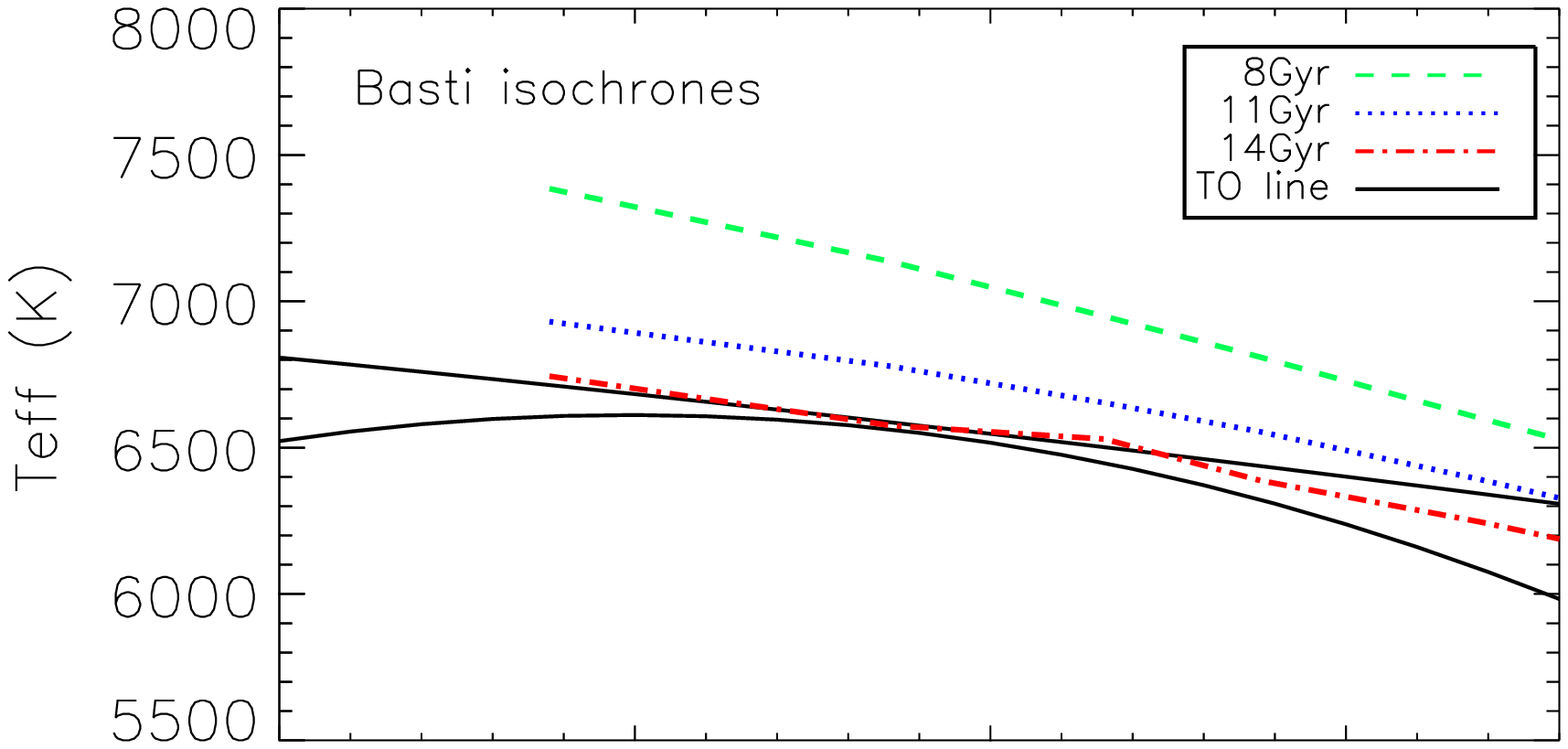}
\includegraphics[scale=0.45,angle=90]{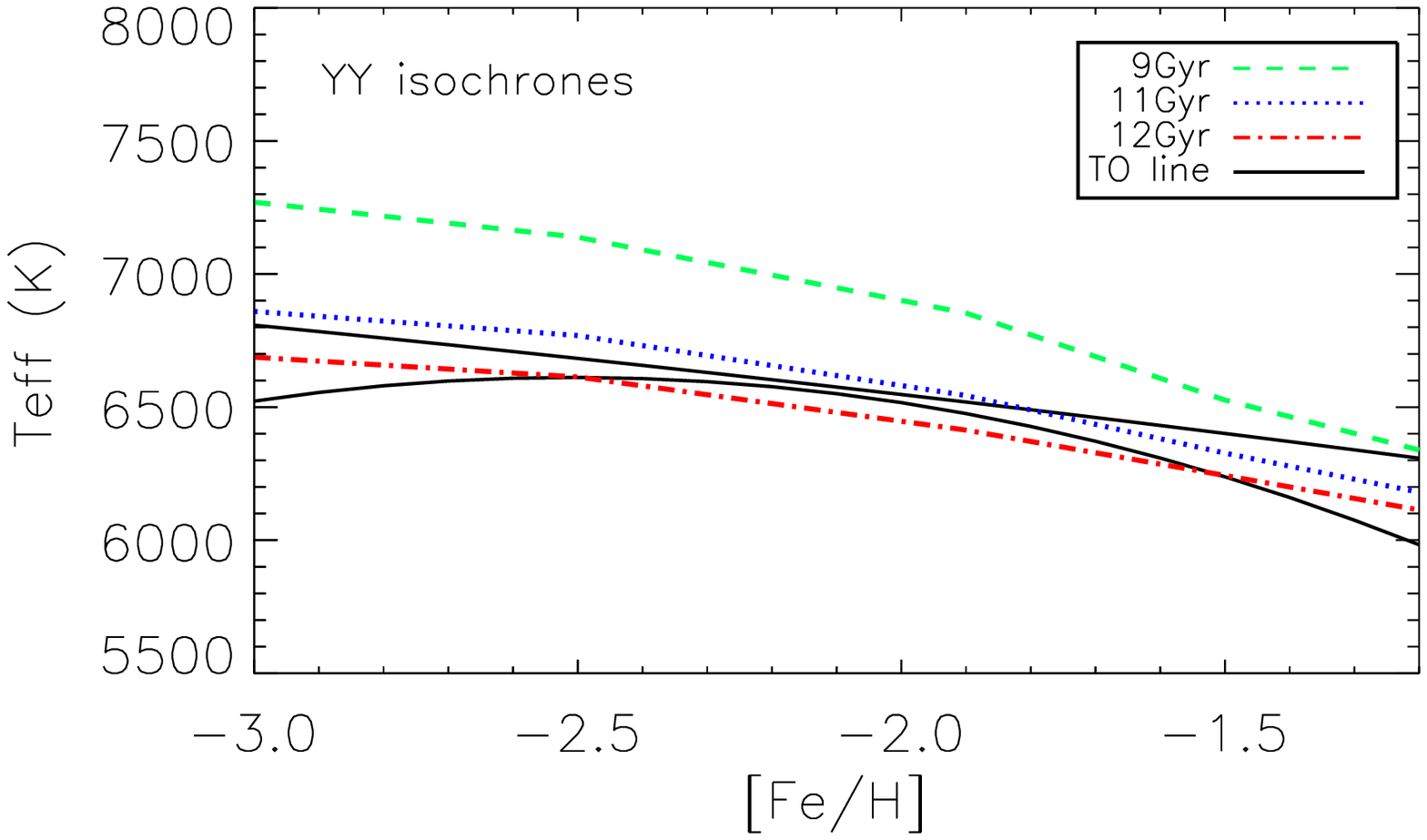}
\vspace{0.4cm}
\caption{Solid line: MSTO temperature range as a function of metallicity of the \textit{Gblue} sample. Color lines: different TO-isochrones of BASTI  (top) and Y$^2$ (bottom).}
\label{fehteff_YY}
\end{figure}

The turn-off temperature as a function of metallicity for the \textit{G blue} sample and TO-isochrones\footnote{TO-temperature of isochrones at different metallicities} is  illustrated in \fig{fehteff_iso}. The MSTO  temperature  range  when considering the bootstrap errors  (see Table~\ref{tab_msto}) as a function of metallicity is also plotted. The upper panel shows the nodiff-isochrones and the lower panel the diff-isochrones. 
There is a need for different ages to reproduce the field main sequence turn-off temperature  when  considering or ignoring diffusion. When diffusion is inactive, the 14-16~Gyr isochrones are those that agree with the  turn-off temperature. On the other hand, when  diffusion is active,  those isochrones that coincide with the MSTO temperature are the 10-12~Gyr ones. There is a mean difference of 4~Gyr in the absolute age due to diffusion, as proposed by \cite{salaris_idea}.  The results obtained for ages from the nodiff-isochrones agree better with those of \cite{unavane96} and \cite{SN96}.

\subsubsection{Comparison with other isochrones}\label{yale}

As a comparison with the results obtained with GARSTEC models, we plotted the turn-off temperature of the  BaSTI \citep{basti, basti2} and the Y$^2$   \citep[Yonsei-Yale, ][]{YY04} isochrones as a function of metallicity along our \fehteff~ diagram in \fig{fehteff_YY}.  We chose BaSTI isochrones as an example of models  computed without  atomic diffusion and the Y$^2$  as an example for models that include diffusion. For both sets of isochrones, we also considered only those with $\alpha-$enhanced chemical composition.
When using  BaSTI models, as illustrated in the upper panel of \fig{fehteff_YY},  isochrones with  more than 14 Gyr are needed to reproduce the turn-off temperature in the metallicity range under study.  This  is consistent with the results obtained for the GARSTEC nodiff-isochrones. 
On the other hand, the lower panel of  \fig{fehteff_YY} shows how Y$^2$ isochrones with ages of 9-12 Gyr are those that reproduce the MSTO temperature, which  agrees with the results we obtained using diff-isochrones. 

It is important to discuss the implications of this test for the debate about the effectiveness of  atomic diffusion mentioned above. If there is no settling, or  if the settling is completely inhibited by turbulent mechanisms, then these stars would be older than the Universe. This is a strong argument in favor of diffusion acting in population II field stars. Only isochrones with  atomic diffusion  yield  realistic ages of 10-12 Gyr for the dominant population of field stars in the Galactic halo.  Moreover, these results for the ages agree with those of GC, as is shown in \sect{gc}.

\begin{figure}[!ht]
\centering
\includegraphics[scale=0.45,angle=90]{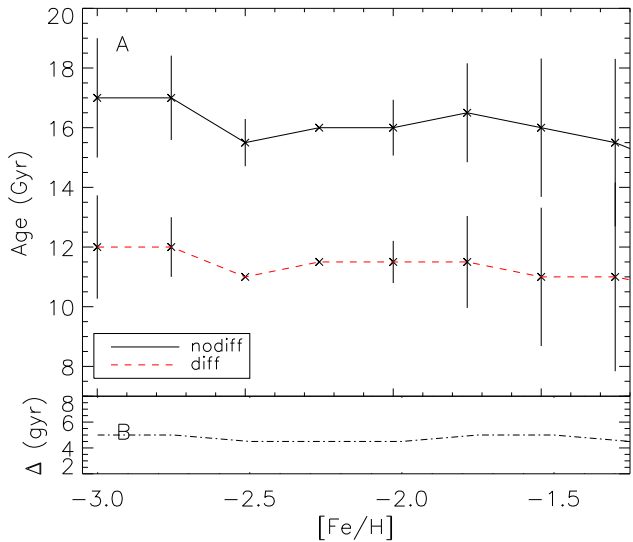}
\caption{Upper panel: The age as a function of metallicity. The red line corresponds to the ages obtained using GARSTEC isochrones with diffusion and the black line to those without diffusion. Lower panel: absolute difference between both ages. }
\label{age_dinodi} 
\end{figure}

\subsection{The age as a function of metallicity}\label{age_met}

The final ages  are listed in \tabl{tab_msto}. The fourth column corresponds to the values obtained with the isochrones that consider diffusion and the fifth column  gives the results from isochrones without atomic diffusion.  The errors considered here are only those connected with the errors obtained for the TO-detection in \sect{msto}, which are indicated in \tabl{tab_msto} as $\sigma_{\mathrm{TO}}$.  We determined the errors for ages considering the turn-off temperature of $\mathrm{T_{eff,TO}} + \sigma_{\mathrm{TO}}$ and $\mathrm{T_{eff,TO}} - \sigma_{\mathrm{TO}}$.  

The ages  as a function of metallicity are displayed in the upper panel of  \fig{age_dinodi}.    In the  full metallicity domain, the ages obtained with nodiff-models are larger than those obtained with diff-isochrones. The difference between ages obtained with and without diffusion as a function of metallicity is displayed in the lower panel of  \fig{age_dinodi}, where the mean difference  is about 4~Gyr. We are aware that this value  could vary when using other stellar evolutionary models.  The upper panel of \fig{age_dinodi} shows that when considering the errors, no real trend is found for age as a function of metallicity. Although the oldest ages are found on the metal-poor side of the metallicity range, the decrease in age towards higher metallicities is negligible when considering the error bars. The reason for the errors increasing at the borders of the metallicity domain is mostly that the uncertainties in  the MSTO detection also increase at the borders. For the age determination, we also considered the uncertainties of 0.25 dex in the metallicity measurements. To do this, we looked at the turn-off temperature of the isochrones with $ \mathrm{\feh} = \mathrm{\feh_{pop} -0.25}$  and $\mathrm{\feh} = \mathrm{\feh_{pop}} +0.25$, where $\mathrm{\feh_{pop}}$ corresponds to the metallicity of the population. \\

 \begin{figure}[!ht]
\centering
\includegraphics[scale=0.45,angle=90]{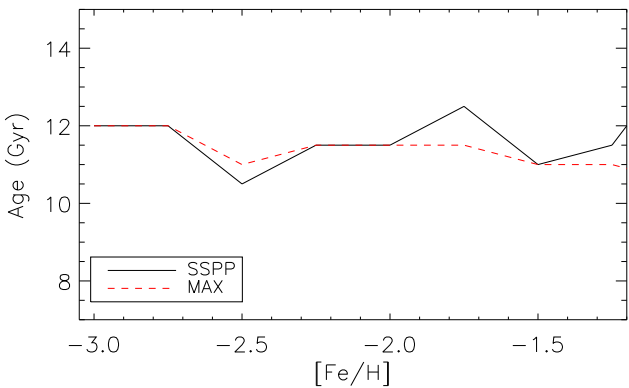}
\vspace{0.5cm}
\caption{Ages for the MSTO temperature as a function of metallicity. Black line indicates the results obtained using SSPP parameters \citep{lee08_1} and the red line those from \maxe~ parameters \citep{max}. The offset  between SSPP and \maxe~ metallicities produce negligible effects in the final age determination. }
\label{age_max_sspp} 
\end{figure}

\noindent  We do not find any significant gradient in age and metallicity. This implies that there is a dominant halo population that is coeval and has a small  scatter in age of less than 2 Gyr.  A similar result was found by \cite{SN06}, who also see that ages of the halo field stars do not have a significant age-metallicity relation.  
As a consistency check, we determined the age for the \textit{G blue} sample when considering the atmosphere parameters provided by the SSPP.   This is shown in \fig{age_max_sspp}, where the age as a function of \feh~ for the MSTO temperature obtained from  SSPP parameters and from the \maxe~ parameters is plotted.  Both results are obtained using diff-isochrones and agree very well. There is a small difference of 0.5~Gyr, with no trend in  age as a function of \feh.  This agreement shows the robustness of the MSTO temperature provided by the dominant halo field population from the SDSS survey. 

The question of whether an age-metallicity relation exists in the Galactic halo is a longstanding problem. Much more effort has been dedicated to finding an age-metallicity relation for the GCs than for the field. The general belief today from GCs is that the outer halo of the Galaxy was formed slowly, rather chaotically, with the Galaxy accreting material over several Gyr, while the inner halo was formed rapidly via a gravitational collapse. An age-metallicity relation for this scenario is expected to have a narrow age spread at low metallicities (representing the rapid collapse of the inner halo) and a broad spread at high metallicities (representing the merging of younger extragalactic systems). The limit between narrow and broad spreads in ages is, however, not set \citep[e.g.][]{SW02, de_angeli05, marin_franch09, Dotter10}.

Finally, we recall that inner halo stars are $\alpha$-enhanced population II stars, meaning that they were formed in an environment where supernova type II dominated the interstellar medium enrichment, thus the formation timescale for these stars was very short \citep[less than 1 Gyr;][]{helmi08}. The age difference for the population II stars of our sample of 1-2 Gyr agrees with this short formation timescale scenario.  As discussed in \cite{gilmore}, studies of the correlations between stellar kinematics and metallicities show that the stars that formed during the dissipational collapse are those with $\mathrm{\feh} \lesssim -1.5$, and their age range is set by the $\alpha-$enhanced chemical composition. 
Our results  tell us that we are mainly observing inner halo stars, which were formed rapidly on a short timescale during the collapse.  \cite{SN10} find a large number of $\alpha-$poor ([$\alpha/\mathrm{Fe]} \sim + 0.2$) stars in the Galactic halo, which are interpreted as  members of an accreted population. These stars have $\mathrm{\feh} > -1.6$. 
It would be interesting to continue to study the kinematic and [$\alpha/$Fe] properties of these stars in our sample.

\begin{figure*}[t]
\begin{center}
\includegraphics[angle=90, scale=0.65]{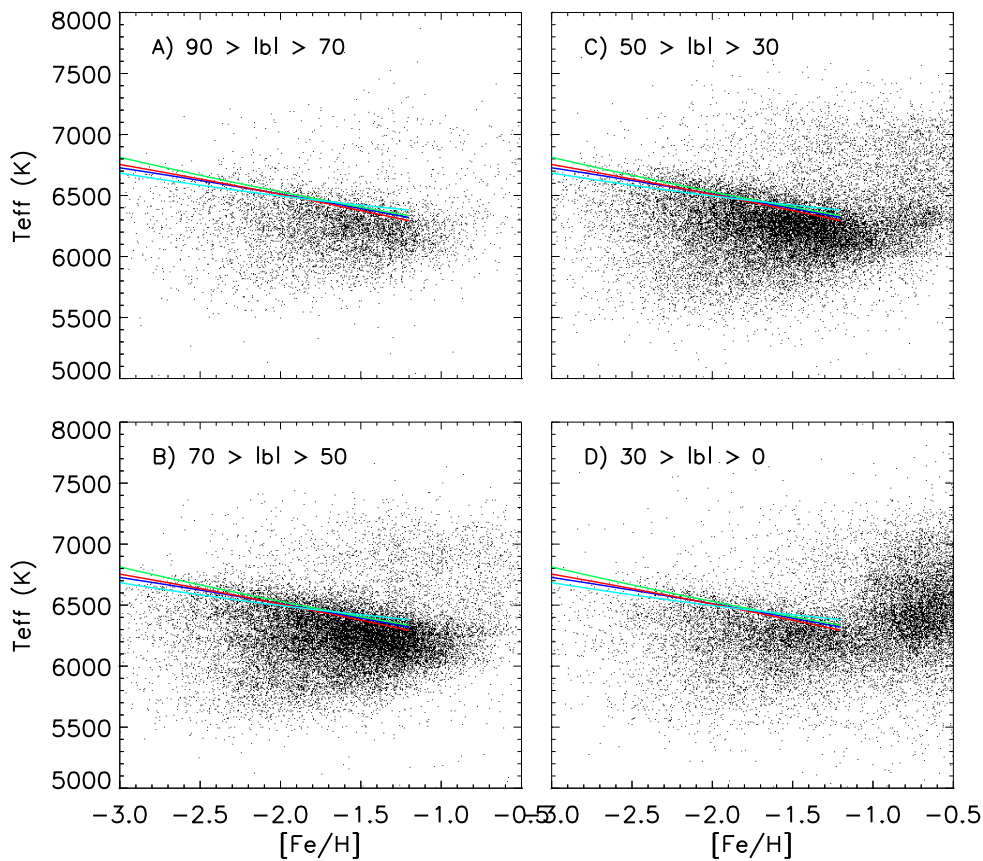}
\caption{\fehteff~ diagrams for four different Galactic latitudes. Color lines correspond to the TO temperature as a function of metallicity, with blue for $90 > |b| > 70 \degrees$, green for  $70 > |b| > 50 \degrees$, red for $50 > |b| > 30 \degrees$ and cyan for $30 > |b| > 0 \degrees$.}
\label{fehteff_sky}
\end{center}
\end{figure*}

\subsection{Connection to the disk}

In addition to halo stars, the SDSS/SEGUE survey contains a large number of disk stars, which usually have higher metallicities than halo stars. It is interesting to study how the halo and the disk components are connected in the \fehteff~ diagram, to see whether it is possible to detect the turn-off, and to determine the age of that population.

For this study we  looked at different Galactic latitudes of the \textit{G blue} stellar sample described in \sect{atm_params}, whose Galactic positions are known from the SDSS database. 
The \fehteff~ diagrams for  four groups of different Galactic latitudes are illustrated in Fig.~\ref{fehteff_sky}, where the regions are indicated in each panel. The parameters used for this study are those of the SSPP pipeline \citep{lee08_1} because the \maxe~ parameters are focused on metal-poor halo stars and not metal-rich disk stars.  The cut-off due to the main-sequence turn-off can be estimated using the Sobel Kernel for stars with $\mathrm{\feh} < -1.3$, which is plotted with different colors depending on the Galactic latitude (see caption of the figure).

We now  discuss the shape of the \fehteff~ diagram as a function of latitude.  Let $N_{+}$ be the number of stars with  \feh $> -1.0$ and $N_{-}$ the number of stars with \feh $< -1.0$. Near the Galactic poles ($90 > |b| > 70 \degrees$) there is a lack of stars with high metallicities, with a ratio of $N_{+} / N_{-} = 0.052$. The number of high-metallicity stars increases with decreasing Galactic latitude, whereas close to the Galactic equator ($30 > |b| > 0 \degrees$) we  find more metal-rich than metal-poor stars, with $N_{+} / N_{-} =1.24$. This fits with the picture  that the Galactic pole contains mainly halo stars -- e.g, metal-poor stars -- and the equator contains mainly disk stars -- e.g, metal-rich stars \citep[e.g.][]{deJong10}.  In the metallicity distributions of \cite{AP06},  the intersection between halo and disk  is at  $\mathrm{\feh} = -1.0$, which agrees with what we observe in the \fehteff~ diagrams of \fig{fehteff_sky}.

We can also see an agreement of the MSTO temperature between the different regions. This means that the turn-off stars in the halo have the same age, regardless of their location in the Galaxy.
The disk stars at higher metallicities, however,  can have bluer colors  than the halo stars, and they do not present an  abrupt cut-off in the temperature distribution  as the halo does. This could be interpreted as  the stars in the disk not representing a dominant population, and  the presence of hotter stars tells us that there are considerably more younger  stars in the disk than in the halo.


\section{Field stars and globular clusters}\label{gc}

Absolute ages of GCs are known with better confidence than the ages of field stars.  Therefore, we used a sample of GCs  observed with the SDSS $ugriz$ filters to consolidate our results. The sample is described in \sect{sample_gc}. We discuss observational evidence of a common nature between cluster and field stars relating their colors  in \sect{obs_evidences}. Based on this evidence we use the cluster ages to constrain the ages of the turn-off halo stars in \sect{ages_gc}. 
\subsection{Globular clusters sample}\label{sample_gc}

The $ugriz$ photometry of a sample of 17 clusters  \citep[hereafter An08]{An08} can be found on the SDSS web pages\footnote{http://www.sdss.org/dr7/products/value\_added/ \\ anjohnson08\_clusterphotometry.html}. The advantage in using  this photometry is that we avoid color transformations, which can introduce systematic errors \citep{WS99, An09}.  We excluded NGC 2419, Pal 3, Pal 14, Pal 4 and NGC 7006 from the sample of  An08 clusters  because they are too faint for SDSS cameras to detect those stars with magnitudes at the turn-off and main-sequence.   In addition, we excluded the cluster M71 because it is more metal-rich ($\mathrm{\feh} = -0.81$) than the metallicity domain used in this work.  The resulting 11 clusters are summarized in Table~\ref{tab_gc}.  The metallicities  listed in the third column are those of An08. That work used the  scale of \cite{KI03} for all the clusters except those with an asterisk, which use that of \cite{harris96}. The fourth column of \tabl{tab_gc} contains  the color of the turn-off (explained below) and the the fifth column contains the reddening value from \cite{schlegel_map}. The last column has the value of the $r$ magnitude of the turn-off taken from the fiducial sequences of \cite{An08}.

\begin{table}[b]

\begin{center}
\begin{tabular}{c  c | c c c c}
\hline
Cluster &  & [Fe/H] & $(g-r)_\mathrm{edge}$ & $E(B-V)$ & $r_{\mathrm{TO}}$\\
NGC & other & \\
\hline

4147 &  & -1.79 & (0.22) & 0.03 & 20.25\\ 
5024 & M53 &  -1.99$^\ast$ & 0.18 & 0.02 & 20.25 \\
5053 & &  -2.41 &0.17 & 0.02 & 19.95\\
5272 & M3 & -1.50 & 0.23 & 0.01 & 18.95\\
5466 & & -2.22$^\ast$ &0.19 & 0.02 & 19.65\\
5904 & M5 & -1.26 & (0.28) & 0.04 & 18.35\\ 
6205 & M13 & -1.60 &0.25 & 0.02 & 18.45\\
6341 & M92 & --2.38 &(0.22) & 0.02 & 18.45\\
7078 & M15 & -2.42 & 0.25 & 0.03 & 19.05\\
7089 & M2 & -1.56 &0.26 & 0.04 & 19.35\\
& Pal 5 &  -1.41$^\ast$ &(0.34) & 0.06 & 20.65\\

\hline

\end{tabular}
\end{center}
\caption{Globular clusters properties taken from \cite{An08} Metallicities adopted are those of \cite{KI03}, except those with asterisk  which have metallicity adopted by \cite{harris96}. The blue edge of the color distribution calculated using the Sobel Kernel technique is indicated in the forth column. Values in parenthesis are the adopted turn-off color taken from the fiducial sequence of the cluster \citep{An08}. The reddening from \citet{schlegel_map} are indicated in the fifth column and the $r$ magnitude of the  turn-off of the fiducial sequence is listed in the last column. }
\label{tab_gc}
\end{table}%

\subsection{Observational evidence}\label{obs_evidences}

Without any previous  assumption concerning Galaxy structure and formation or stellar evolution theory, we present in the following pieces of observational evidence that halo cluster and field stars may have a similar nature. To analyze the stars comparatively, we first considered their magnitudes corrected by extinction. For field stars, the magnitudes of the \textit{G blue} sample were already de-reddened using the $E(B-V)$ coefficients of \cite{schlegel_map}. For the clusters, the values of the extinction coefficient $E(B-V)$ are indicated in \tabl{tab_gc} \citep[taken from An08, which are also those of][]{schlegel_map}. To transform these coefficients to $ugr$ filters, we used the ratios of selective extinction $R_u$, $R_g $, and $R_r$ of \cite{McCall}.

\begin{figure}[t]
\centering
\hspace{0.3cm}
\includegraphics[scale=0.4,angle=90]{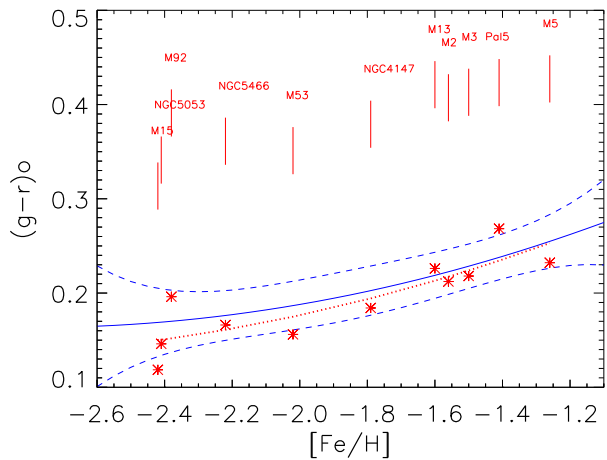}
\vspace{0.3cm}
\caption{The solid line corresponds to the blue egde as a function of metallicity for the \textit{G blue} sample of field stars and the dashed line the bootstrap error. Red asterisks indicate the blue edge of the dereddened $(g-r)$ color and metallicity of the clusters of \cite{An08}. Their names are indicated at the top of the diagram.}
\label{col_feh}
\end{figure}

In \fig{col_feh} we illustrate a comparison between the bluest color of the globular cluster and the field stars as a function of metallicity. To determine the blue edge for clusters and field, we adapted the Sobel Kernel technique (\sect{TO_TRGB}) to find the edge in the color distribution.   The value of the  $(g-r)$ color is indicated in the fourth column of Table~\ref{tab_gc} as $(g-r)_\mathrm{edge}$ (not corrected for reddening).  The numbers in parenthesis in \tabl{tab_gc} represent the bluest point of the cluster fiducial sequence taken from the Tables 9-28 of An08 because for these cases we were unable to find a good estimate of the blue edge using the Sobel Kernel technique.

In \fig{col_feh} we observe how the bluest color as a function of metallicity of the field stars agrees well with that of the GCs within the errors, with metal-poor stars having a bluer turn-off color than metal-rich ones.  The trend of the blue edge of the field stars  corresponds to the quadratic polynomial 

\begin{equation}
(g-r)_{\mathrm{edge,field}} = 0.467 + 0.217\cdot \mathrm{[Fe/H]} + 0.039 \cdot \mathrm{[Fe/H]}^2.
\end{equation}
It is very similar to the trend of the GCs, which is
\begin{equation}
(g-r)_{\mathrm{edge,GC}} = 0.471 + 0.217 \cdot \mathrm{[Fe/H]} + 0.035 \cdot \mathrm{[Fe/H]}^2.
\end{equation}

\noindent The coefficients of the parabola are very similar and have an offset of 0.04 mag at the zero point,  which is less than the bootstrapping errors obtained for the field stars.   This suggests that the age obtained for the field stars should also agree with that of the  clusters.

\begin{figure}[t]
\centering
\includegraphics[scale=0.47,angle=90]{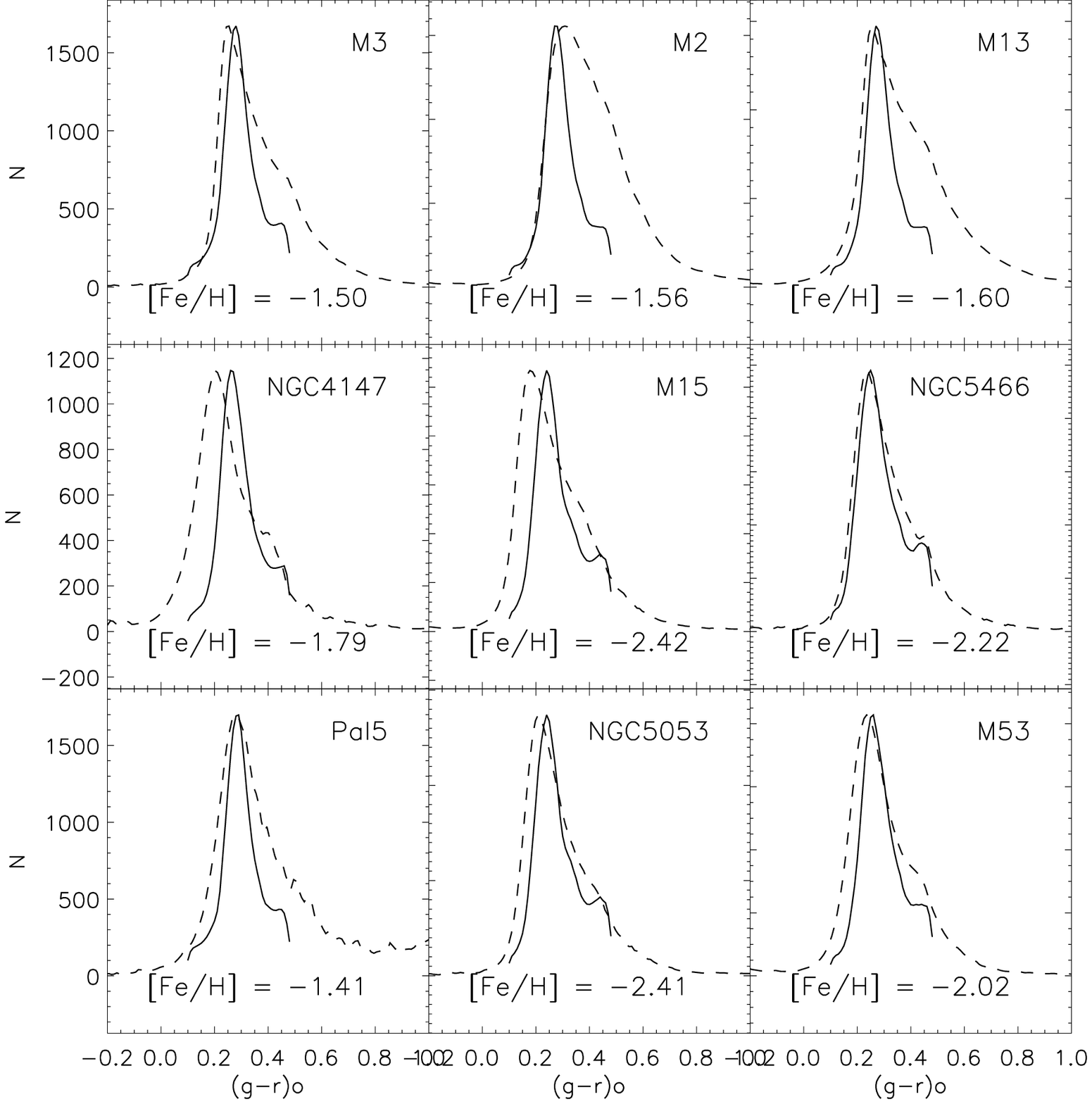}
\caption{Color distribution of the GCs of \tabl{tab_gc} with dashed line. The distribution for the field stars with the same metallicity as the cluster is plotted with a solid line.}
\label{col_dis}
\end{figure}

\subsection{Color distribution}

We selected the stars with metallicities of $(\mathrm{\feh_{GC}}- 0.15, \mathrm{\feh_{GC}}+ 0.15)$, where $\mathrm{\feh_{GC}}$ is the metallicity of the cluster, and calculated the probability function given by Eq.~(\ref{eq:phi}) of the color distribution. The same probability function was determined for the color distribution of the clusters. Comparisons for the 11 GCs are displayed in \fig{col_dis}. The clusters and the field population at the cluster metallicity usually do not have the same number of stars, therefore we scaled the distribution to compare them better.  In \fig{col_dis}, the metallicity and the name of the clusters are indicated in each panel.

\begin{figure}[t]
\centering
\includegraphics[scale=0.5,angle=90]{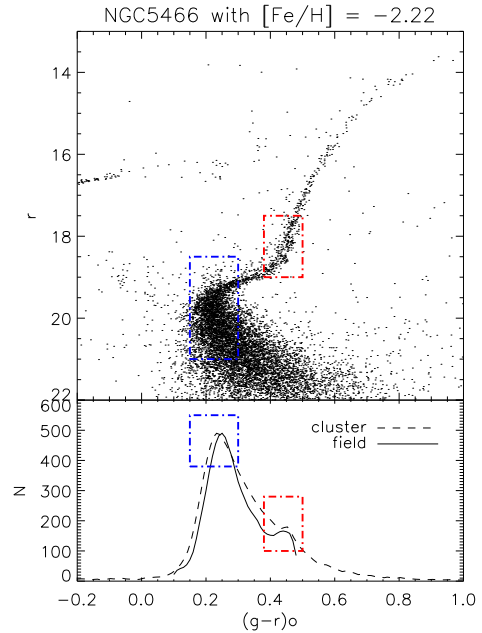}
\caption{Upper panel: NGC5466 color-magnitude diagram. Lower panel: color distribution of the cluster (dashed line) and the field of cluster metallicity  (solid line), smoothed with the probability function given by Eq.~\ref{eq:phi}.}
\label{ex_distr}
\end{figure}

To understand the color distributions better, we took the cluster NGC5466 as an example and plotted it separately in \fig{ex_distr}. The first similarity seen in Figs.~\ref{col_dis} and \ref{ex_distr} is the shape of the distributions. This is  especially notable at the turn-off, shown in \fig{ex_distr}. The color of the maximum number of stars in the GC agrees with the respective field stellar population, as can also be observed  in \fig{col_dis}.  There are exceptions, like NGC4147, where the color distribution of the cluster is bluer than for the field. Even in this case the shape of both distributions agrees. 

 Towards the red side of the color distribution,  the number of stars decreases in both distributions. The decreasing slope for the field is steeper than for the clusters.  This is mainly from selection effects. The  \textit{G blue} sample is biased towards  turn-off stars. In addition, the field halo dwarfs observed by SDSS usually have magnitudes of $r \sim 19 - 20$ mag \citep{SEGUE}. Since the SDSS survey cannot observe stars below $r \sim 22$ mag \citep{dr7}, the samples from this survey  have only a  few distant halo stars in the main-sequence.  Figure~\ref{col_dis} shows us that those clusters with a luminous turn-off (see \tabl{tab_gc})  have a well-populated region on the red side of the color distribution. 
The opposite is observed for clusters with  a less luminous TO. Finally,  clusters with strong contamination of field stars,  such as M2, M3, and Pal5,  have very noisy CMD (see the color-magnitude diagrams  in An08) and show a very populated red region. 

\begin{figure}[t]
\centering
\includegraphics[scale=0.45,angle=90]{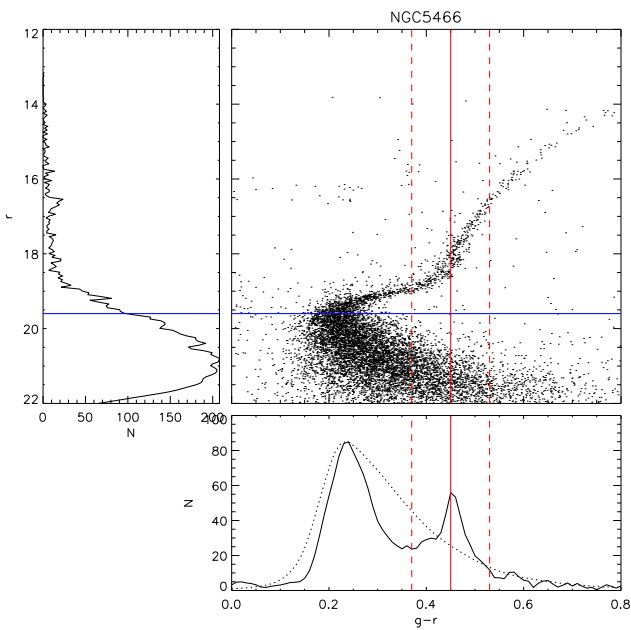}
\caption{CMD of NGC 5466. The turn-off magnitude of the fiducial sequence \citep{An08} is represented by the blue horizontal line. Lower panel:  color distributions of dwarfs (dotted line) and giants (solid line) of the cluster, separated with the limit of the TO magnitude. The distribution of the giants show a second peak, with its maximum represented by the red vertical line. Dashed lines indicate the width of the peak.  Left panel: luminosity function of the cluster.}
\label{bump}
\end{figure}

 Another  feature of the red side of the distribution  is a minor peak at $(g-r)_o \sim 0.45$  seen in each panel of \fig{col_dis}. These stars  are enclosed by the red dashed box in \fig{ex_distr} and can be found in the globular clusters quite easily. We can separate the dwarfs from the giants of the cluster by using the fiducial TO magnitude, as illustrated  in \fig{bump}. We calculated the color distributions for the dwarfs and giants separately, as shown in the lower panel of \fig{bump}. Only the giants present peaks at red colors. This peak is the result of the accumulation of stars at the base of the red giant branch \citep[hereafter BGB]{iben67}.  The high concentration of stars at this color is an optical  effect, because the base of the giant branch is the point where the red giant branch is the steepest and the stars have the same temperature, regardless of their mass and evolutionary stage.  Stars that lie at redder colors than the BGB are more massive and evolve faster.  Therefore, few of these stars are found in the cluster, and this second agglomeration of stars then decreases again to the red end of the color distribution.  Only the temperature of the BGB is independent of mass. The luminosity  varies for stars with different masses. This is the reason there is no accumulation of stars at the BGB  in the luminosity function plotted on the left side of \fig{bump}.

\begin{figure*}[t]
\centering
\includegraphics[scale=0.55,angle=90]{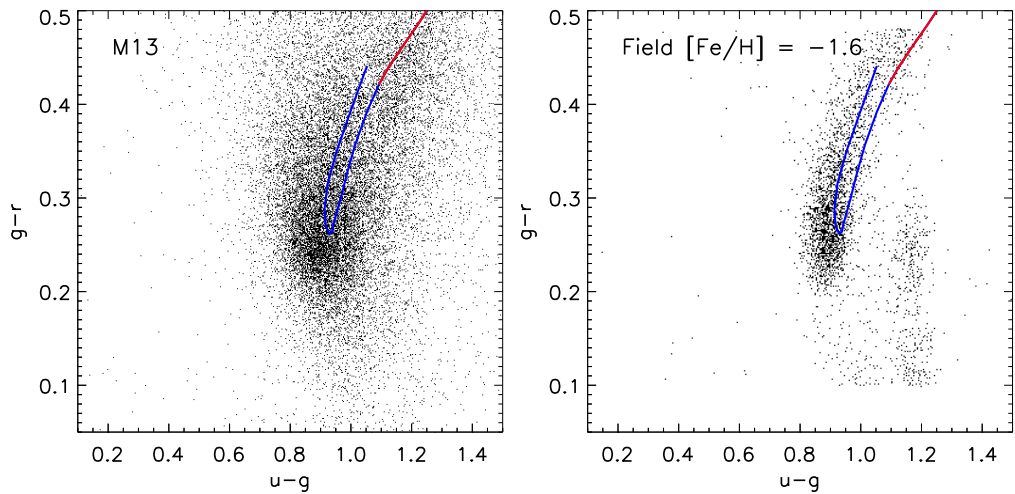}
\caption{Left panel: $(u-g) - (g-r)$ color-color diagram of M13. Right panel: color-color diagram of the field stars with $\mathrm{ \feh} = -1.6$, according to M13 metallicity value.  Blue line illustrates the fiducial sequence of the cluster of \cite{An08}. Red color corresponds to the stars at the base of the red giant branch. Red giant branch stars in the field are expected from  our color constraint.}
\label{colcol_bump}
\end{figure*}

 Some  contamination with BGB is expected in the \textit{G blue} sample, because their colors lie in the color constraint of this sample (see \sect{atm_params}). An example is shown in \fig{colcol_bump}, where  color-color diagrams of the cluster M13 and the  field stars with M13 metallicity are illustrated in the left and right panels, respectively. The blue line corresponds to the fiducial sequence. The base of the red giant branch is indicated in red. It is possible to see how  red giant branch stars are included in the $G blue$ sample due to our color constraint (see \sect{atm_params}).

\subsubsection{Consequences}\label{implications}

The excellent agreement of the blue edge as a function of metallicity of clusters and field stars seen in the \feh-color diagram of \fig{col_feh}  supports the idea that we are finding the main-sequence turn-off for a population in the Galactic halo. This statement is reinforced when looking at the color distribution of the field and GCs.  The majority of the stars are located close to the turn-off for both field and cluster stellar populations, and the rate of decrease towards blue and red colors is similar. In addition, the minor peak seen on the red side of the color distribution corresponds to a sample of red giant branch stars of the same population. 

A consequence of these results is the possible common origin of GC and halo field stars, although our color distributions do not reveal the formation scenario. In particular, we can see here that field and cluster stars in our study may have been formed at the same time and  in similar chemical environments  \citep{first_stars}. 

Another consequence, probably more extreme, is that a significant number of stars in the Galactic halo were formed in GCs.  Although simulations of the dynamical evolution of GCs is still not completely clear \citep{dercole08, baumgardt08, carretta10}, it is accepted that they lose stars. Observations of extratidal structures of GCs \citep{ martinez_delgado04, law10, clusters17} in the Galactic halo confirm this result.  But the question of how many field halo stars come from disrupted GCs is still under debate \citep[e.g.][]{yong08, boley09, martell10, carretta10}.  Recently, \cite{clusters17}  detected tidal features using the same data set as this study. This means that this particular sample of  clusters is being evaporated into the Galactic halo. Although the extra-tidal stars of these clusters contribute a small fraction of the entire field halo population, they are being removed  through tidal stripping. This fits well with the similar color distributions seen here.  

The case of NGC~4147  is worth mentioning . It shows a well-defined color distribution but shifted to bluer colors with respect to the field distribution at the NGC~4147 metallicity. The results of \cite{Dotter10} reveal that this cluster has an age of 12.75~Gyr, which puts it in the frame of the normal old inner halo clusters. However,  \cite{Meissner06}  obtained an age of 9~Gyr for this cluster, which would explain its blue shift.

Finally we want to recall our statement in \sect{age_met} that this sample of field stars is mainly formed from one coeval dominating population. Globular clusters are composed of stars with the same age and metallicity\footnote{Although GCs are composed by multiple populations \citep[e.g.][]{gratton04}, as tracers of the Galaxy they can be treated as single stellar populations.}. Therefore, the observations discussed here are consistent with this picture.  Moreover, we will see in the next section that these  clusters are coeval with the field stars within 1-2~Gyr.

\begin{table*}[t]

\begin{center}
\begin{tabular}{|c  c | c c  |c c  |c c|}
\hline
Cluster &  & An09  & & SW02  && D10 & \\ 
NGC & other & \feh &Age (Gyr) &\feh&Age (Gyr)& \feh& Age (Gyr) \\
\hline

4147 & $\cdots$ &  $\cdots$  & $\cdots$  &$\cdots$ & $\cdots$ & $-1.7$ & $12.75 \pm 0.75$ \\
5024 & M53 &  $\cdots$ &  $\cdots$  & $\cdots$  &$\cdots$ &  $-2.0$ & $13.25 \pm 0.5$ \\
5053 & $\cdots$ & $\cdots $ & $\cdots$ & $-1.98$ &     $10.8 \pm 0.9$  & $-2.4$ &     $13.5 \pm 0.75$     \\
5272 & M3 & $-1.5$ & $13.3 \pm 1.4$ & $-1.33 $ & $11.3 \pm 0.7$ &$-1.6$& $12.5 \pm 0.5$ \\
5466 & $\cdots$ & $\cdots$ & $\cdots$ & $-2.13$ & $12.2 \pm 0.9$ & $-2.1$ & $13.00 \pm 0.75$\\
5904 & M5 & $-1.26$ & $12.2 \pm 1.3$ & $-1.12$ & $10.9 \pm 1.1$ & $-1.3$& $12.25 \pm 0.75$\\
6205 & M13 & $-1.6$ & $14.3 \pm 1.1$ & $-1.33$ & $11.9 \pm 1.1$ & $-1.6 $ & $13.00 \pm 0.5$\\
6341 & M92 & $-2.38$ & $14.4 \pm 0.9$ &$ -2.10 $& $12.3 \pm 0.9$ & $-2.4$ & $13.25\pm 1$\\
7078 & M15 & $-2.42$ & $13.9 \pm 2.5$ & $-2.02$ & $11.7 \pm 0.8$ & $-2.4$ & $13.25 \pm 1.0$ \\
7089 & M2 & $\cdots$ & $\cdots$ & $\cdots$ & $\cdots$ & $-1.6$ & $12.50 \pm 0.75$\\
$\cdots$ & Pal 5 & $\cdots$ & $\cdots$ & $-1.47$ & $10.0 \pm 1.4$  & $\cdots $& $\cdots$ \\

\hline
\end{tabular}
\caption{Ages of the sample of globular clusters determined in three different works: \citet[An09]{An09}, \citet[SW02]{SW02} and \citet[D10]{Dotter10}. For the metallicity and method adopted by the authors see text.}
\label{age_gc}
\end{center} 
\end{table*}%

\subsection{Globular cluster ages}\label{ages_gc}

From the observational evidence and implications discussed above, it is  expected that the 11 GCs studied here have ages according to the dominating halo population.  We  discussed in \sect{age} the behavior of the absolute ages obtained from the main-sequence turn-off. In particular where we saw  how the diffusion parameter can reduce the age by 4 Gyr.  The age of GCs is much better constrained, and we can use it to  consolidate the ages for the field stars. We considered the ages of \citet[hereafter An09]{An09}, \citet[hereafter SW02]{SW02}, and \citet[hereafter D10]{Dotter10}. These are shown in \tabl{age_gc}, including the metallicity scale adopted by the authors. The second block of the table indicates the values of An09, the third block the values of SW02, and the last one those of D10.

Although the three works use nodiff-isochrones to determine the ages, the methods employed differ between them. An09 and D10 fit the isochrones to the color-magnitude diagram near the turn-off, making their results being affected by atomic diffusion. The method used by SW02 to determine the ages, on the other hand, consists in a combination of the $\Delta V(\mathrm{TO-HB})$ and $\Delta{(V-I)}$ methods \citep{chaboyer92a}. This approach is affected very weakly by the assumption of gravitational settling \citep[for details  of this age-dating method, see][]{Meissner06}.

\begin{figure}[t]
\centering
\includegraphics[scale=0.47,angle=90]{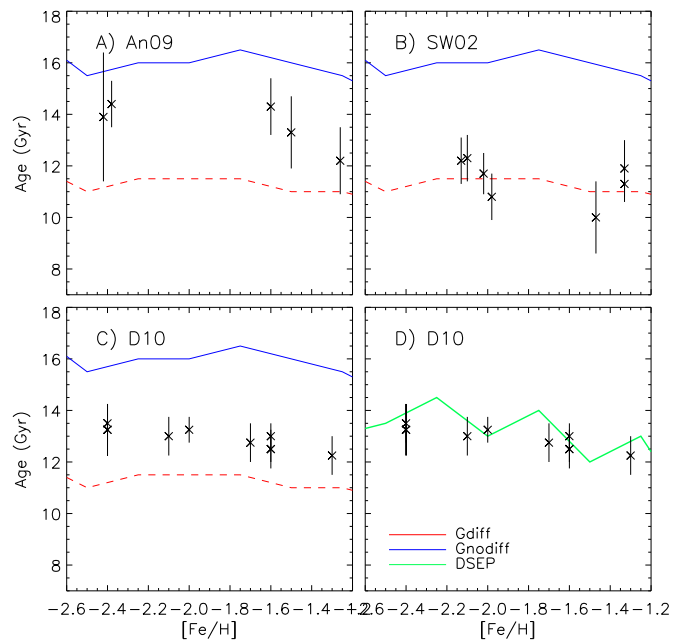}
\caption{Panel A: The age as a function of \feh~ for the cluster sample of \cite{An09} represented by asterisks. The red dashed and blue solid lines are the field ages obtained using isochrones with and without diffusion, respectively (see \sect{age}). Panel B: Same as panel A but with asterisks representing the ages and \feh~ of \cite{SW02}. Panel C: Same as panels A and B, plotting this time the ages and \feh~ of \cite{Dotter10} with asterisks. Panel D: Again D10 clusters but  compared with field ages using DSEP isochrones.}
\label{age_met_gc}
\end{figure}

The comparison between the ages of the GCs and the field stars is illustrated in \fig{age_met_gc}. We considered the ages of the field stars obtained using diff- and nodiff-isochrones (see \sect{age}). Panel A shows the comparison with the An09 results. The absolute ages of the clusters agree with our results obtained with the nodiff-isochrones, as expected. We can see, however, that the trend in age as a function of metallicity is different for the clusters and the field stars, but we cannot conclude that this is related to a physical difference in the chemical enrichment of field and cluster stars.

In panel B of \fig{age_met_gc} we find the age and metallicity of the SW02 clusters.  In this case, we can see a similar trend between the age and the metallicity for the clusters and field stars.  In the SW02 analysis,  55 clusters were included, and there was also no age gradient at  $\mathrm{\feh} < -1.2$.    Interestingly,  our final absolute age using diff-isochrones agrees  with that of the clusters. This is further argument in favor of using isochrones with atomic diffusion for the age determination of field population II stars. 

Panel C compares the D10 results with our owns.  The absolute values of the ages are  different because the isochrones employed by D10 do not inhibit diffusion completely.  They inhibit settling only in the outer regions of the radiative core. In the outermost  region, diffusion is completely stopped. In the region below, diffusion has been linearly ramped from zero to the full effect \citep{Dotter08}. This partial diffusion  produces a MSTO temperature that is hotter than a fully diff-isochrone and colder than a nodiff-isochrone \citep[see e.g.][]{chaboyer01}.  We downloaded a set of isochrones from the DSEP database\footnote{http://stellar.dartmouth.edu/$\sim$models/} and determined the age of the field stars as described in \sect{age}. The results are illustrated in panel D of \fig{age_met_gc}, where we also plotted the D10 ages for the GCs.  The agreement between clusters and field ages is now excellent, as expected.

It is important to discuss here the importance of using the color of the turn-off for determining the ages of GCs. Not only atomic diffusion affects the effective temperature at the turn-off  significantly, but other uncertainties, such as the mixing length parameter, metallicity scale, and color calibrations, can affect the absolute ages of globular clusters by 4 Gyr as well \citep{BV01}. This is one of the main reasons for GC ages being better determined using the brightness of the turn-off \citep[see e.g. ][]{Meissner06}.  The absolute ages obtained for the halo field stars by \cite{SN06} of 13 Gyr agree with the usual GCs ages, even when their isochrones do not employ atomic diffusion, is because the color transformations of the \cite{BV01} isochrones are adjusted specifically to match the colors of the old metal-poor GCs.

In summary, we have adopted three different scales for ages and metallicities of 11 clusters to calibrate the age of the Galactic halo field population.  We saw how they show different results, where An09 yield higher ages than D10.  This is because An09 isochrones do not consider diffusion, while D10 isochrones partially do.  We saw  excellent agreement with the D10 scale and the field ages when the same isochrones were used.  
The SW02 ages employed a different method from An09, D10, and our own. It has the advantage of  very weakly depending on atomic diffusion and ad hoc adjustments in the color transformations, making it a more robust calibrator of absolute ages for the Galactic halo. It is encouraging to obtain  good agreement in the absolute scale given by the diff-isochrones and the GCs, especially because we have avoid color transformations in our procedure. This reinforces our statement that the inner halo has an age of 10-12 Gyr, where field and cluster stars were formed at the same time.

\section{Summary and concluding remarks}\label{fin}

We determined the age of the dominant Galactic halo population by using  a sample of  field stars from the Sloan Digital Sky Survey \citep[SDSS,][]{sdss}. 

The first main task was to determine the stellar atmosphere parameters from the SDSS sample. We used an efficient method \citep[\maxe,][]{max}  to obtain accurate parameters in a timely manner.  
Based on these parameters,  we determined the turn-off temperature for the Galactic halo as a function of metallicity, from which we could  estimate the age of the halo  stars. The turn-off was identified using a robust edge-detector kernel.  We studied the final value of the turn-off when using the atmosphere parameters recovered by the SEGUE Stellar Parameter Pipeline \citep{lee08_1}. We showed that it agreed with the value obtained using  our own atmosphere parameters.  In addition, we could see the excellent agreement of the turn-off color obtained for the SDSS spectroscopic sample with that of the  \cite{SN06} photometric sample. This  implies that the edge produced by the temperature of the main sequence turn-off is a definite feature of the Milky Way halo, independent of the data set and method employed for finding it. 

For the age determination, we used the temperature of the turn-off and the metallicity of the stellar populations to find the isochrones of that particular TO temperature and metallicity. We explored the effect of atomic diffusion  in the resulting  ages and found an absolute difference of 4~Gyr for ages obtained ignoring or including atomic diffusion in the stellar models.  The age was additionally determined using Y$^2$ \citep{YY04} and BaSTI \citep{basti} isochrones, as an example of isochrones with and without diffusion, respectively. We could see that the results obtained using BaSTI models agreed better with those obtained from the nodiff-isochrones, while Y$^2$ results agreed better with those obtained from diff-isochrones.    In this way, we could explain why the previous ages obtained with this method \citep{unavane96, SN96} are greater than our own. 
 Given the current debate about the efficiency with which atomic diffusion acts in stellar interiors, our results show a strong argument against fully inhibited diffusion in metal-poor halo stars. If fully inhibited diffusion was the case, these stars  would be older than  the Universe.  Based on this argument, we could give an age for the field stars of the Galactic halo of 10-12~Gyr. 

The relation between the age and the metallicity of the different stellar populations does not show a gradient. Our sample of stars could be a representation of the inner halo, which could be a product of a rapid star formation scenario, probably during the collapse phase of the proto-Galactic cloud.  Additionally, a breakdown in the relation at $\mathrm{ \feh}  = -1.6$ could be related to a stronger contamination of younger thick disk and outer halo stars.

 For the consolidation of the absolute age for the Galactic  inner halo, we considered the ages of GCs, for which the parameters are better determined than for field stars. We found observational evidence that cluster and field stars share a similar history and are composed of a dominant population. For 11 clusters of \cite{An08}, which were  also observed with the SDSS telescopes, we compared the colors  to those of the field stars. We could see that both color distributions agreed especially in the turn-off color as a function of metallicity.  The agreement between the ages for the field and those obtained by \cite{SW02} for GCs serves as a further argument that the majority of inner halo stars are 10-12~Gyr old. \\  

\noindent An old dominating population  for the inner halo suggests that the Milky Way halo  could have formed rapidly during a collapse of the proto-Galactic gas, where star formation took place  10-12~Gyr ago. This scenario  agrees with conclusions from GC studies \citep{sarajedini97, SW02, de_angeli05, Dotter10}, where the metal-poor clusters are coeval. Moreover, the absolute ages obtained for the clusters by \cite{SW02} are also 10-12~Gyr, which implies that GCs and field stars are coeval as well. 

In our \fehteff~ diagrams it is possible to observe a significant number of stars with bluer colors than the color cut-off of the main sequence turn-off. They could be BHB stars, white dwarfs, blue stragglers of metal-poor stars, but in any case bluer than the dominant population. They were also noticed by  \cite{unavane96} and \citet[and references therein]{SN06} and are referred as blue metal-poor stars \citep{BMP, preston00}. An interesting explanation for them is that they were formed in  small external galaxies and have been accreted later on to the Milky Way halo, supporting the current hierarchical galaxy formation scenario. These galaxies  experience a different star formation history to the Milky Way and therefore can be younger than the dominating population of  field stars.  Further analysis of their kinematics and chemical abundances is needed to prove this scenario. These blue stars and the existence of a dominant population of halo stars suggest that the two rivaling formation scenarios of the Galactic halo from \cite{eggen62} and \cite{searle_zinn} actually complete, in a composed manner, the picture of how the Milky Way might have formed. These scenarios combined suggest that part of the halo has collapsed rapidly, while the other part  has been populated through collisions and mergers  between the satellite galaxies and our Milky Way.

\begin{acknowledgements}
This work is part of the Ph.D. thesis of Paula Jofr\'e and has been funded by the IMPRS fellowship. We wish to thank M. Salaris and P. Demarque for their contribution through helpful discussions. We also acknowledge the contribution of L. Sbordone in building our grid of atmosphere models for this analysis. P. Jofr\'e thanks especially to P. Das, T. M\"adler, and R. Yates for their careful reading of the manuscript. Finally, the authors are sincerely thankful the constructive suggestions made by the referee.

    Funding for the SDSS and SDSS-II has been provided by the Alfred P. Sloan Foundation, the Participating Institutions, the National Science Foundation, the U.S. Department of Energy, the National Aeronautics and Space Administration, the Japanese Monbukagakusho, the Max Planck Society, and the Higher Education Funding Council for England. The SDSS Web Site is http://www.sdss.org/.

    The SDSS is managed by the Astrophysical Research Consortium for the Participating Institutions. The Participating Institutions are the American Museum of Natural History, Astrophysical Institute Potsdam, University of Basel, University of Cambridge, Case Western Reserve University, University of Chicago, Drexel University, Fermilab, the Institute for Advanced Study, the Japan Participation Group, Johns Hopkins University, the Joint Institute for Nuclear Astrophysics, the Kavli Institute for Particle Astrophysics and Cosmology, the Korean Scientist Group, the Chinese Academy of Sciences (LAMOST), Los Alamos National Laboratory, the Max-Planck-Institute for Astronomy (MPIA), the Max-Planck-Institute for Astrophysics (MPA), New Mexico State University, Ohio State University, University of Pittsburgh, University of Portsmouth, Princeton University, the United States Naval Observatory, and the University of Washington.

\end{acknowledgements}
\bibliographystyle{aa} 
\bibliography{refs_ageMW}
\clearpage
\end{document}